\documentclass[conference]{IEEEtran}
\pdfoutput=1
\IEEEoverridecommandlockouts

\usepackage[T1]{fontenc}

\usepackage{orcidlink}
\usepackage{float}
\usepackage{amssymb}
\usepackage{todonotes}
\usepackage{hyperref}
\usepackage{graphicx}
\usepackage[most]{tcolorbox}
\usepackage{tikz}
\usetikzlibrary{calc,scopes,arrows.meta,arrows,decorations,shapes,intersections,patterns}
\usetikzlibrary{fit,fadings,automata,positioning, angles,quotes,babel,shadows,shadows.blur}
\usepackage{tikz-3dplot}
\usepackage{xifthen}
\usepackage{amssymb}

\tikzset{fit margins/.style={/tikz/afit/.cd,#1,
    /tikz/.cd,
    inner xsep=\pgfkeysvalueof{/tikz/afit/left}+\pgfkeysvalueof{/tikz/afit/right},
    inner ysep=\pgfkeysvalueof{/tikz/afit/top}+\pgfkeysvalueof{/tikz/afit/bottom},
    xshift=-\pgfkeysvalueof{/tikz/afit/left}+\pgfkeysvalueof{/tikz/afit/right},
    yshift=-\pgfkeysvalueof{/tikz/afit/bottom}+\pgfkeysvalueof{/tikz/afit/top}},
    afit/.cd,left/.initial=2pt,right/.initial=2pt,bottom/.initial=2pt,top/.initial=2pt}

\definecolor{BlauHigh}{RGB}{96,150,180}
\definecolor{BlauMid}{RGB}{147,191,207}
\definecolor{BlauLow}{RGB}{189,205,214}
\definecolor{BlauDark}{RGB}{49,74,93}
\definecolor{Beige}{RGB}{238,233,218}
\definecolor{GreenHigh}{RGB}{119, 191, 163}
\definecolor{GreenMid}{RGB}{152, 201, 163} %
\definecolor{GreenLow}{RGB}{164, 195, 178}
\definecolor{RedHigh}{RGB}{237, 106, 94}
\definecolor{RedMid}{RGB}{255, 142, 114}
\definecolor{RedLow}{RGB}{255, 175, 135}
\definecolor{rotx}{RGB}{128,29,26}
\definecolor{DeepGreen}{RGB}{0,64,32}
\definecolor{ClaySoft}{RGB}{225,200,185}
\definecolor{BlueFog}{RGB}{204,213,222}
\definecolor{SoftPeach}{RGB}{255,193,150}

\def\blockWidth{2.5cm}
\def\blockHeight{1cm}

\tikzset{
    shadow/.style={drop shadow={shadow xshift=.5ex, shadow yshift=-.5ex}}, 
    shadedBlauHigh/.style={top color=BlauHigh, bottom color=BlauHigh!90, draw=black},
    artifactLow/.style={draw=none, fill=BlauLow,shadow, minimum width=\blockWidth, minimum height=\blockHeight,align=center,text=white},
    artifactMid/.style={draw=none, fill=BlauMid,shadow, minimum width=\blockWidth, minimum height=\blockHeight, align=center,text=white},
    artifactHigh/.style={draw=none, fill=BlauHigh,shadow, minimum width=\blockWidth, minimum height=\blockHeight,align=center,text=white},
    concept/.style={draw=white,ultra thick, fill=BlauDark,shadow, minimum width=\blockWidth*1.5, minimum height=\blockHeight*1.5,align=center, font=\bfseries\large,text=white},
    context/.style={draw=none, fill=white,shadow, minimum width=\blockWidth, minimum height=\blockHeight,align=center,rounded corners=3pt, draw=BlauDark},
    evidence/.style={draw=BlauDark, fill=white,shadow, align=center,circle=2cm, minimum height=\blockHeight*.8,minimum width=\blockWidth*.8},
    areaGreen/.style={draw=black, fill=GreenLow, shadow,}, 
    areaRed/.style={draw=RedHigh, fill=RedHigh!15, shadow},
    areaBeige/.style={left color=Beige, right color=Beige!35, shadow, draw=black},
    areaBlue/.style={draw=BlauDark, fill=BlauDark!15, shadow},
    buffer/.style={shape border rotate=270,regular polygon,regular polygon sides=3,minimum height=1.5cm,fill=Beige},
    myarrow/.style={-{Triangle[width=7pt, length=5pt]},thick,BlauDark},
    aggregation/.style={{Diamond[open, width=8pt, length=16pt,fill=white]}-,thick},
    myarrowWhite/.style={-{Triangle[width=7pt, length=5pt]},thick,white},
    myarrowWhiteopen/.style={-{Triangle[open,width=7pt, length=5pt]},thick,white},
    myarrowopen/.style={-{Triangle[open, width=7pt, length=5pt]},thick,BlauDark},
    basicBlock/.style={fill=white,shadow, minimum width=\blockWidth, minimum height=\blockHeight,align=center, draw=BlauDark},
    ontologyElement/.style={fill=SoftPeach,shadow, draw=black,minimum width=\blockWidth, minimum height=\blockHeight,align=center,text=black},
    goal/.style={fill=white,shadow, minimum width=\blockWidth, minimum height=\blockHeight,align=center, draw=BlauDark},
    strategy/.style={fill=white,shadow, minimum width=\blockWidth, minimum height=\blockHeight,align=center, draw=BlauDark, trapezium,trapezium left angle =70, trapezium right angle = 110},
    connectorLabel/.style={pos=.5, font=\normalsize,text=black}
    }

\pgfdeclarelayer{background3}
\pgfdeclarelayer{background2}
\pgfdeclarelayer{background1}
\pgfdeclarelayer{main}
\pgfdeclarelayer{foreground}
\pgfsetlayers{background3,background2,background1,main,foreground}

\usepackage{scalerel,stackengine,amsmath}

\usepackage{cleveref}
\usepackage{color}

\usepackage[normalem]{ulem}
\newcommand\reduline{\bgroup\markoverwith{\textcolor{blue!80}{\rule[-0.8ex]{2pt}{.5pt}}}\ULon}

\newcommand{\ArtifactLegend}{\raisebox{-.9mm}{\tikz{\filldraw[draw=white,ultra thick,shadow,  fill=BlauDark,text=white,line width=.5pt](0,0) rectangle ++ (8mm,3.4mm) node[pos=.5,white] {};}}\hspace*{0cm}}
\newcommand{\OntologyLegend}{\raisebox{-.9mm}{\tikz{\filldraw[draw=black,thick, shadow,fill=SoftPeach,text=black,line width=.5pt](0,0) rectangle ++ (8mm,3.4mm) node[pos=.5,white] {};}}\hspace*{0cm}}
\newcommand{\GoalLegend}{\raisebox{-.9mm}{\tikz{\filldraw[shadow,draw=BlauDark,thick, fill=white,text=black,line width=.5pt](0,0) rectangle ++ (8mm,3.4mm) node[pos=.5,white] {};}}\hspace*{0.1cm}}
\newcommand{\SolutionLegend}{\raisebox{-.9mm}{\tikz{\node[shadow,circle,draw=BlauDark,thick,fill=white,text=black,line width=.5pt,minimum size=4mm,inner sep=0pt] {};}}\hspace*{0.1cm}}
\newcommand{\ContextLegend}{\raisebox{-.9mm}{\tikz{\filldraw[shadow,draw=BlauDark, fill=white, text=black,line width=.5pt, rounded corners=3pt](0,0) rectangle ++ (8mm,3.4mm) node[pos=.5,white] {};}}\hspace*{0.1cm}}
\newcommand{\StrategyLegend}{\raisebox{-1.8mm}{\tikz{\filldraw[draw=BlauDark,thick, shadow,fill=white,text=black,line width=.5pt](0,0) -- ++ (2mm,3.4mm) -- ++ (6mm,0mm) -- ++ (-2mm,-3.4mm) -- ++ (-6mm,0mm) node[pos=.5,white] {};}}\hspace*{0.1cm}}

\newcommand{\AggregationLegend}{\tikz{\draw[line width=0.5pt, draw=black, -{Diamond[open, width=5pt, length=10pt, fill=white]}] (0,0) -- ++(6mm,0mm);}}
\newcommand{\SatisfyLegend}{\raisebox{-.4mm}{\tikz{\draw[line width=0.5pt, draw=rotx, dashed,-{Triangle[width=7pt, length=5pt]}] (0,0) -- ++(6mm,0mm);}}}

\newcommand{\MacrostructReq}{\raisebox{0.15ex}{$\vcenter{\hbox{\includegraphics[width=0.35cm]{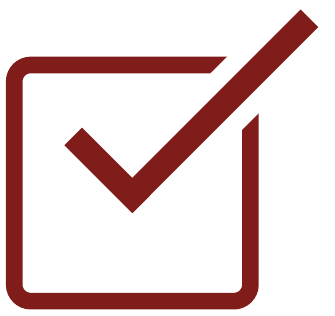}}}$}}
\newcommand{\MicrostructReq}{\raisebox{0.15ex}{$\vcenter{\hbox{\includegraphics[width=0.35cm]{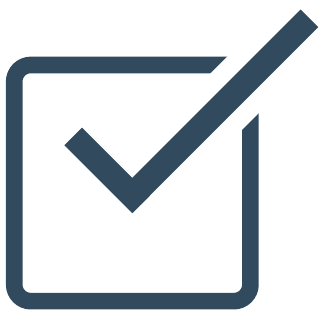}}}$}}
\newcommand{\AdditionalReq}{\raisebox{0.15ex}{$\vcenter{\hbox{\includegraphics[width=0.35cm]{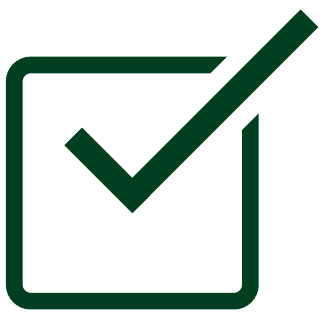}}}$}}

\newcommand{\MicroReq}[1]{\hfill \textbf{\textcolor{BlauDark}{$\blacktriangleright$~R#1}}}
\newcommand{\MacroReq}[1]{\hfill \textbf{\textcolor{rotx}{$\blacktriangleright$~R#1}}}
\newcommand{\AddReq}[1]{\hfill \textbf{\textcolor{DeepGreen}{$\blacktriangleright$~R#1}}}

\newcommand{\MicroReqText}[1]{\textbf{\textcolor{BlauDark}{$\blacktriangleright$~R#1}}}
\newcommand{\MacroReqText}[1]{\textbf{\textcolor{rotx}{$\blacktriangleright$~R#1}}}
\newcommand{\AddReqText}[1]{\textbf{\textcolor{DeepGreen}{$\blacktriangleright$~R#1}}}

\newcommand{\fon}[1]{\fontfamily{#1}\selectfont}

\definecolor{rotx}{RGB}{128,29,26}
\definecolor{BlauDark}{RGB}{49,74,93}
\definecolor{DeepGreen}{RGB}{0,64,32}

\crefformat{figure}{Fig.~#2#1#3}

\usepackage[style=ieee,
backend=biber,
minnames=1,
maxcitenames=6,
maxbibnames=5,
doi=true,
isbn=false,
url=false,
natbib=true,
date=year
]{biblatex}
\usepackage{xpatch}
\usepackage{xstring}

\DeclareSourcemap{
	\maps[datatype=bibtex]{
		\map{
			\step[fieldsource=langid, match=\regexp{\A(n)?((swiss)?german|austrian)\Z}, final]
			\step[fieldset=language, fieldvalue={(in German)}]
		}
		\map{
			\step[fieldsource=langid, match=pinyin, final]
			\step[fieldset=language, fieldvalue={(in Chinese)}]
		}
		\map{
			\step[fieldsource=langid, match=japanese, final]
			\step[fieldset=language, fieldvalue={(in Japanese)}]
		}
		\map{
			\step[fieldsource=langid, match=ko, final]
			\step[fieldset=language, fieldvalue={(in Korean)}]
		}
	}
}

\xpatchbibdriver{inproceedings}					 	
	{\usebibmacro{publisher+location+date}}			
	{\IfSubStr{\strfield{booktitle}}{\strfield{year}}{\usebibmacro{publisher+location}}{\usebibmacro{publisher+location+date}}} 
	{} 
	{\typeout{There was an error patching biblatex-ieee (specifically, ieee.bbx's @inproceedings driver)}} 

\newbibmacro*{publisher+location}{%
	\printlist{location}%
	\iflistundef{publisher}
	{\setunit*{\addcomma\space}}
	{\setunit*{\addcolon\space}}%
	\printlist{publisher}%
	\newunit}

\xpatchbibdriver{online}
{\printtext[parens]{\usebibmacro{date}}}
	{\iffieldundef{year}
		{}
		{\printtext[parens]{\usebibmacro{date}}}}
	{}
	{\typeout{There was an error patching biblatex-ieee (specifically, ieee.bbx's @online driver)}}

\DeclareSourcemap{
	\maps[datatype=biber]{
		\map{
			\step[fieldsource=note, final]
			\step[fieldset=addendum, origfieldval, final]
			\step[fieldset=note, null]
		}
	}
}

\DeclareBibliographyDriver{standard}{%
	\usebibmacro{bibindex}%
	\usebibmacro{begentry}%
	\usebibmacro{maintitle+title}%
	\setunit{\addcomma\addspace}%
	\usebibmacro{publisher+type+number}%
	\setunit{\labelnamepunct}\newblock
	\IfSubStr{\strfield{number}}{\strfield{year}}{}{\usebibmacro{date}} 
	\newunit\newblock
	\usebibmacro{finentry}%
}

\newbibmacro*{publisher+type+number}{%
	\printtext{%
		\printlist{publisher}
		\iflistundef{publisher}
		{\space}{}%
		\iffieldundef{type}
		{Standard}
		{\printfield{type}}
		\printfield{number}
	}%
}

\xpatchbibdriver{report}					 	
	{\usebibmacro{date}}			
	{\IfSubStr{\strfield{number}}{\strfield{year}}{}{\usebibmacro{date}}} 
	{} 
	{\typeout{There was an error patching biblatex-ieee (specifically, ieee.bbx's @report driver)}} 

\DeclareBibliographyDriver{software}{%
	\usebibmacro{bibindex}%
	\usebibmacro{begentry}%
	\usebibmacro{maintitle+title}%
	\setunit{\addcomma\addspace}%
	\usebibmacro{version+year}%
	\setunit{\labelnamepunct}\newblock
	\usebibmacro{author}%
	\setunit{\addcomma\addspace}%
	\newunit\newblock
	\usebibmacro{url+urldate}%
	\newunit\newblock
	\usebibmacro{finentry}%
}

\newbibmacro*{version+year}{%
	\printtext{%
		\iffieldundef{version}{\iffieldundef{year}{}{\printfield{year}}}{\printfield{version}}
		\printfield{number}
		\printlist{publisher}
		\iflistundef{publisher}{\space}{}%
	}%
}

\DeclareSourcemap{
	\maps[datatype=bibtex]{
		\map[overwrite, foreach={booktitle,journaltitle,eventtitle,series,publisher,institution,type}]{
			\step[fieldsource=\regexp{$MAPLOOP}, match={XXX}, replace={Acad. Serbe Sci. Arts Glas Cl. Sci. Tech.}]
			\step[fieldsource=\regexp{$MAPLOOP}, match={XXX}, replace={Acta Acust.}]
			\step[fieldsource=\regexp{$MAPLOOP}, match={XXX}, replace={Acta Astron. (Poland)}]
			\step[fieldsource=\regexp{$MAPLOOP}, match={XXX}, replace={Acta Astron. Sin. (China)}]
			\step[fieldsource=\regexp{$MAPLOOP}, match={XXX}, replace={Acta Astronaut. (U.K.)}]
			\step[fieldsource=\regexp{$MAPLOOP}, match={XXX}, replace={Acta Astrophys. Sin. (China)}]
			\step[fieldsource=\regexp{$MAPLOOP}, match={XXX}, replace={Acta Autom. Sin.}]
			\step[fieldsource=\regexp{$MAPLOOP}, match={XXX}, replace={Acta Cienc. Indica Math.}]
			\step[fieldsource=\regexp{$MAPLOOP}, match={XXX}, replace={Acta Cienc. Indica Phys.}]
			\step[fieldsource=\regexp{$MAPLOOP}, match={XXX}, replace={Acta Crystallogr. A, Found. Crystallogr.}]
			\step[fieldsource=\regexp{$MAPLOOP}, match={XXX}, replace={Acta Crystallogr. B, Struct. Sci.}]
			\step[fieldsource=\regexp{$MAPLOOP}, match={XXX}, replace={Acta Crystallogr. C, Cryst. Struct. Commun.}]
			\step[fieldsource=\regexp{$MAPLOOP}, match={XXX}, replace={Acta Electron. (France)}]
			\step[fieldsource=\regexp{$MAPLOOP}, match={XXX}, replace={Acta Cyberno}]
			\step[fieldsource=\regexp{$MAPLOOP}, match={XXX}, replace={Acta Electron. Sin. (China)}]
			\step[fieldsource=\regexp{$MAPLOOP}, match={XXX}, replace={Acta Geod. Geophys. Montan. Hung.}]
			\step[fieldsource=\regexp{$MAPLOOP}, match={XXX}, replace={Acta Geophys. Pol.}]
			\step[fieldsource=\regexp{$MAPLOOP}, match={XXX}, replace={Acta Geophys. Sin. (China)}]
			\step[fieldsource=\regexp{$MAPLOOP}, match={XXX}, replace={Acta Geophys. Sin. (USA)}]
			\step[fieldsource=\regexp{$MAPLOOP}, match={XXX}, replace={Acta Metall.}]
			\step[fieldsource=\regexp{$MAPLOOP}, match={XXX}, replace={Acta Mex. Cienc. Tecnol.}]
			\step[fieldsource=\regexp{$MAPLOOP}, match={XXX}, replace={Acta Phys. Hung.}]
			\step[fieldsource=\regexp{$MAPLOOP}, match={XXX}, replace={Acta Phys. Pol. A}]
			\step[fieldsource=\regexp{$MAPLOOP}, match={XXX}, replace={Acta Phys. Pol. B}]
			\step[fieldsource=\regexp{$MAPLOOP}, match={XXX}, replace={Acta Phys. Sin.}]
			\step[fieldsource=\regexp{$MAPLOOP}, match={XXX}, replace={Acta Phys. Slovaca}]
			\step[fieldsource=\regexp{$MAPLOOP}, match={XXX}, replace={Acta Politec. Mex.}]
			\step[fieldsource=\regexp{$MAPLOOP}, match={XXX}, replace={Acta Polytech. Scand. Appl. Phys. Ser.}]
			\step[fieldsource=\regexp{$MAPLOOP}, match={XXX}, replace={Acta Polytech. Scand. Chem. Technol.}]
			\step[fieldsource=\regexp{$MAPLOOP}, match={XXX}, replace={Metall. Ser.}]
			\step[fieldsource=\regexp{$MAPLOOP}, match={XXX}, replace={Acta Polytech. Scand. Electr. Eng. Ser.}]
			\step[fieldsource=\regexp{$MAPLOOP}, match={XXX}, replace={Acta Polytech. Scand. Math. Comput. Sci. Ser.}]
			\step[fieldsource=\regexp{$MAPLOOP}, match={XXX}, replace={Acta Polytech. Scand. Mech. Eng. Ser.}]
			\step[fieldsource=\regexp{$MAPLOOP}, match={XXX}, replace={Acta Seismol. Sin.}]
			\step[fieldsource=\regexp{$MAPLOOP}, match={XXX}, replace={Acta Tech. Acad. Sci. Hung.}]
			\step[fieldsource=\regexp{$MAPLOOP}, match={XXX}, replace={Acta Tech. CSAV}]
			\step[fieldsource=\regexp{$MAPLOOP}, match={XXX}, replace={Acustica}]
			\step[fieldsource=\regexp{$MAPLOOP}, match={XXX}, replace={(AEU) Arch. Elektr. Ubertragung}]
			\step[fieldsource=\regexp{$MAPLOOP}, match={XXX}, replace={Akust. Zh.}]
			\step[fieldsource=\regexp{$MAPLOOP}, match={XXX}, replace={Algorithmica}]
			\step[fieldsource=\regexp{$MAPLOOP}, match={XXX}, replace={Alta Freq.}]
			\step[fieldsource=\regexp{$MAPLOOP}, match={XXX}, replace={An. Acad. Bras. Cienc.}]
			\step[fieldsource=\regexp{$MAPLOOP}, match={XXX}, replace={An. Fis.}]
			\step[fieldsource=\regexp{$MAPLOOP}, match={XXX}, replace={An. Mec. Electr.}]
			\step[fieldsource=\regexp{$MAPLOOP}, match={XXX}, replace={Angew. Inform.}]
			\step[fieldsource=\regexp{$MAPLOOP}, match={XXX}, replace={Ann. Inst. Henri Poincare Phys. Theor.}]
			\step[fieldsource=\regexp{$MAPLOOP}, match={XXX}, replace={Ann. Soc. Sci. Brux. I, Sci. Math. Astron. Phys.}]
			\step[fieldsource=\regexp{$MAPLOOP}, match={XXX}, replace={Arch. Elektr. Uebertrag. (AEU)}] 
			\step[fieldsource=\regexp{$MAPLOOP}, match={XXX}, replace={Arch. Elektron. Uebertrag. Tech.}] 
			\step[fieldsource=\regexp{$MAPLOOP}, match={XXX}, replace={Arch. Elektrotech. (Poland)}]
			\step[fieldsource=\regexp{$MAPLOOP}, match={XXX}, replace={Arch. Elektrotech. (Germany)}]
			\step[fieldsource=\regexp{$MAPLOOP}, match={XXX}, replace={Ark. Fys. Semin. Trondheim}]
			\step[fieldsource=\regexp{$MAPLOOP}, match={XXX}, replace={Astrofizika}]
			\step[fieldsource=\regexp{$MAPLOOP}, match={XXX}, replace={Astron. Nachr. (Germany)}]
			\step[fieldsource=\regexp{$MAPLOOP}, match={XXX}, replace={Astron. Tidsskr.}]
			\step[fieldsource=\regexp{$MAPLOOP}, match={XXX}, replace={Astron. Vestn.}]
			\step[fieldsource=\regexp{$MAPLOOP}, match={XXX}, replace={Astron. Zh.}]
			\step[fieldsource=\regexp{$MAPLOOP}, match={XXX}, replace={Atomwirtsch.-Atomtech.}]
			\step[fieldsource=\regexp{$MAPLOOP}, match={XXX}, replace={Atti Accad. Sci. Ist. Bologna CI. Sci. Fis.}]
			\step[fieldsource=\regexp{$MAPLOOP}, match={XXX}, replace={Rend. XIII}]
			\step[fieldsource=\regexp{$MAPLOOP}, match={XXX}, replace={Atti Accad. Sci. Torino I, CI. Sci. Fis. Math. Nat.}]
			\step[fieldsource=\regexp{$MAPLOOP}, match={XXX}, replace={Autom. Strum.}]
			\step[fieldsource=\regexp{$MAPLOOP}, match={XXX}, replace={Autom. Tech. Prax.}]
			\step[fieldsource=\regexp{$MAPLOOP}, match={XXX}, replace={Automatica}]
			\step[fieldsource=\regexp{$MAPLOOP}, match={XXX}, replace={Automatie}]
			\step[fieldsource=\regexp{$MAPLOOP}, match={XXX}, replace={Automatika}]
			\step[fieldsource=\regexp{$MAPLOOP}, match={XXX}, replace={Automatisierungstechnik}]
			\step[fieldsource=\regexp{$MAPLOOP}, match={XXX}, replace={Automatizace}]
			\step[fieldsource=\regexp{$MAPLOOP}, match={XXX}, replace={Automedica}]
			\step[fieldsource=\regexp{$MAPLOOP}, match={XXX}, replace={Avtom. Telemekh.}]
			\step[fieldsource=\regexp{$MAPLOOP}, match={XXX}, replace={Avtom. Vychisl. Tekh.}]
			\step[fieldsource=\regexp{$MAPLOOP}, match={XXX}, replace={Avtomatika}]
			\step[fieldsource=\regexp{$MAPLOOP}, match={XXX}, replace={Avtometriya}]
			\step[fieldsource=\regexp{$MAPLOOP}, match={ATZelektronik worldwide}, replace={ATZelektron. worldw.}]
			\step[fieldsource=\regexp{$MAPLOOP}, match={XXX}, replace={Ber. Bunsenges. Phys. Chem.}]
			\step[fieldsource=\regexp{$MAPLOOP}, match={XXX}, replace={Biofizika}]
			\step[fieldsource=\regexp{$MAPLOOP}, match={XXX}, replace={Biometrika}]
			\step[fieldsource=\regexp{$MAPLOOP}, match={XXX}, replace={Boll. Geofis. Teor. Appl.}]
			\step[fieldsource=\regexp{$MAPLOOP}, match={XXX}, replace={Bull. Acad. Serbe Sci. Arts cl. Sci. Tech.}]
			\step[fieldsource=\regexp{$MAPLOOP}, match={XXX}, replace={Bull. Annu. Soc. Suisse Chronom. Lab. Suisse.}]
			\step[fieldsource=\regexp{$MAPLOOP}, match={XXX}, replace={Rech. Horlog.}]
			\step[fieldsource=\regexp{$MAPLOOP}, match={XXX}, replace={Bull. Cl. Sci. Acad. R. Belg.}]
			\step[fieldsource=\regexp{$MAPLOOP}, match={XXX}, replace={Bull. Dir. Etud. Rech. A}]
			\step[fieldsource=\regexp{$MAPLOOP}, match={XXX}, replace={Bull. Dir. Etud. Rech. B}]
			\step[fieldsource=\regexp{$MAPLOOP}, match={XXX}, replace={Bull. Dir. Etud. Rech. C}]
			\step[fieldsource=\regexp{$MAPLOOP}, match={XXX}, replace={Bull. Liaison Rech. Inform. Autom.}]
			\step[fieldsource=\regexp{$MAPLOOP}, match={XXX}, replace={Bur. Etud. Autom.}]
			\step[fieldsource=\regexp{$MAPLOOP}, match={XXX}, replace={CFI-Ceram. Forum Int.-Ber. Dtsch. Keram. Ges.}]
			\step[fieldsource=\regexp{$MAPLOOP}, match={XXX}, replace={Chem. Scr.}]
			\step[fieldsource=\regexp{$MAPLOOP}, match={XXX}, replace={Ciel Terre}]
			\step[fieldsource=\regexp{$MAPLOOP}, match={XXX}, replace={Cybernetica}]
			\step[fieldsource=\regexp{$MAPLOOP}, match={XXX}, replace={Deut. Hydrogr. Z.}]
			\step[fieldsource=\regexp{$MAPLOOP}, match={XXX}, replace={Dokl. Akad. Nauk SSSR}]
			\step[fieldsource=\regexp{$MAPLOOP}, match={XXX}, replace={Electroacoustique}]
			\step[fieldsource=\regexp{$MAPLOOP}, match={XXX}, replace={Electrochim. Acta}]
			\step[fieldsource=\regexp{$MAPLOOP}, match={XXX}, replace={Electrochim. Metal.}]
			\step[fieldsource=\regexp{$MAPLOOP}, match={XXX}, replace={Elektor Electron.}]
			\step[fieldsource=\regexp{$MAPLOOP}, match={XXX}, replace={Elektr. Bahnen}]
			\step[fieldsource=\regexp{$MAPLOOP}, match={XXX}, replace={Elektr. Energ.-Tech.}]
			\step[fieldsource=\regexp{$MAPLOOP}, match={XXX}, replace={Elektr. Masch.}]
			\step[fieldsource=\regexp{$MAPLOOP}, match={XXX}, replace={Elektr. Stn.}]
			\step[fieldsource=\regexp{$MAPLOOP}, match={XXX}, replace={Elektrichestvo}]
			\step[fieldsource=\regexp{$MAPLOOP}, match={XXX}, replace={Elektrie}]
			\step[fieldsource=\regexp{$MAPLOOP}, match={XXX}, replace={Elektrizitaetswirtschaft}]
			\step[fieldsource=\regexp{$MAPLOOP}, match={XXX}, replace={Elektro}]
			\step[fieldsource=\regexp{$MAPLOOP}, match={XXX}, replace={Elektro-Anz.}]
			\step[fieldsource=\regexp{$MAPLOOP}, match={XXX}, replace={Elektro-Jahr}]
			\step[fieldsource=\regexp{$MAPLOOP}, match={XXX}, replace={Elektrokhimiya}]
			\step[fieldsource=\regexp{$MAPLOOP}, match={XXX}, replace={Elektron}]
			\step[fieldsource=\regexp{$MAPLOOP}, match={XXX}, replace={Elektron Int.}]
			\step[fieldsource=\regexp{$MAPLOOP}, match={XXX}, replace={Elektron. Entwitkl.}]
			\step[fieldsource=\regexp{$MAPLOOP}, match={XXX}, replace={Elektron. Ind}]
			\step[fieldsource=\regexp{$MAPLOOP}, match={XXX}, replace={Elektron. J.}]
			\step[fieldsource=\regexp{$MAPLOOP}, match={XXX}, replace={Elektron. Prax.}]
			\step[fieldsource=\regexp{$MAPLOOP}, match={XXX}, replace={Elektron. Tekh.}]
			\step[fieldsource=\regexp{$MAPLOOP}, match={XXX}, replace={Elektronica}]
			\step[fieldsource=\regexp{$MAPLOOP}, match={XXX}, replace={Elektronik}]
			\step[fieldsource=\regexp{$MAPLOOP}, match={XXX}, replace={Elektronika}]
			\step[fieldsource=\regexp{$MAPLOOP}, match={XXX}, replace={Elektroniker}]
			\step[fieldsource=\regexp{$MAPLOOP}, match={XXX}, replace={Elektronikschau}]
			\step[fieldsource=\regexp{$MAPLOOP}, match={XXX}, replace={Elektrosvyaz}]
			\step[fieldsource=\regexp{$MAPLOOP}, match={XXX}, replace={Elektrotech. Cas.}]
			\step[fieldsource=\regexp{$MAPLOOP}, match={Elektrotechnik und Informationstechnik}, replace={Elektrotech. Inf. Tech.}]
			\step[fieldsource=\regexp{$MAPLOOP}, match={XXX}, replace={Elektrotech. Obz.}]
			\step[fieldsource=\regexp{$MAPLOOP}, match={XXX}, replace={Elektrotechniek}]
			\step[fieldsource=\regexp{$MAPLOOP}, match={XXX}, replace={Elektrotechnik (Czechoslovakia)}]
			\step[fieldsource=\regexp{$MAPLOOP}, match={XXX}, replace={Elektrotechnik (Switzerland)}]
			\step[fieldsource=\regexp{$MAPLOOP}, match={XXX}, replace={Elektrotechnik (Germany)}]
			\step[fieldsource=\regexp{$MAPLOOP}, match={XXX}, replace={Elektrotechnika}]
			\step[fieldsource=\regexp{$MAPLOOP}, match={XXX}, replace={Elektrotehnika, Zagreb}]
			\step[fieldsource=\regexp{$MAPLOOP}, match={XXX}, replace={Elektrotekhnika}]
			\step[fieldsource=\regexp{$MAPLOOP}, match={XXX}, replace={Elektroteknikeren}]
			\step[fieldsource=\regexp{$MAPLOOP}, match={XXX}, replace={Elektrowaerme Int. B.}]
			\step[fieldsource=\regexp{$MAPLOOP}, match={XXX}, replace={Elettrificazione}]
			\step[fieldsource=\regexp{$MAPLOOP}, match={XXX}, replace={Elettron. Oggi}]
			\step[fieldsource=\regexp{$MAPLOOP}, match={XXX}, replace={Elettron. Telecomun.}]
			\step[fieldsource=\regexp{$MAPLOOP}, match={XXX}, replace={Elettrotecnica}]
			\step[fieldsource=\regexp{$MAPLOOP}, match={XXX}, replace={Elek. Med Aktuell Elektron.}]
			\step[fieldsource=\regexp{$MAPLOOP}, match={XXX}, replace={Elteknik}]
			\step[fieldsource=\regexp{$MAPLOOP}, match={XXX}, replace={Energ. Atomtech.}]
			\step[fieldsource=\regexp{$MAPLOOP}, match={XXX}, replace={Energ. Elettr.}]
			\step[fieldsource=\regexp{$MAPLOOP}, match={XXX}, replace={Energetica}]
			\step[fieldsource=\regexp{$MAPLOOP}, match={XXX}, replace={Energetik}]
			\step[fieldsource=\regexp{$MAPLOOP}, match={XXX}, replace={Energetika}]
			\step[fieldsource=\regexp{$MAPLOOP}, match={XXX}, replace={Energetyka}]
			\step[fieldsource=\regexp{$MAPLOOP}, match={XXX}, replace={Energia Nuclear}]
			\step[fieldsource=\regexp{$MAPLOOP}, match={XXX}, replace={Energie Technik (Germany)}]
			\step[fieldsource=\regexp{$MAPLOOP}, match={XXX}, replace={Energie Technik (Switzerland)}]
			\step[fieldsource=\regexp{$MAPLOOP}, match={XXX}, replace={Entropie}]
			\step[fieldsource=\regexp{$MAPLOOP}, match={XXX}, replace={ETZ}]
			\step[fieldsource=\regexp{$MAPLOOP}, match={XXX}, replace={ETZ Arch.}]
			\step[fieldsource=\regexp{$MAPLOOP}, match={XXX}, replace={Feingeraetetechnik}]
			\step[fieldsource=\regexp{$MAPLOOP}, match={XXX}, replace={Feinw. Tech. Messtech.}]
			\step[fieldsource=\regexp{$MAPLOOP}, match={XXX}, replace={Fert. Tech. Betr.}]
			\step[fieldsource=\regexp{$MAPLOOP}, match={XXX}, replace={Fis. Tecnol.}]
			\step[fieldsource=\regexp{$MAPLOOP}, match={XXX}, replace={Fiz. Khim. Obrab. Mater.}]
			\step[fieldsource=\regexp{$MAPLOOP}, match={XXX}, replace={Fiz. Met. Metalloved}]
			\step[fieldsource=\regexp{$MAPLOOP}, match={XXX}, replace={Fiz. Nizk. Temp.}]
			\step[fieldsource=\regexp{$MAPLOOP}, match={XXX}, replace={Fiz. Plazmy}]
			\step[fieldsource=\regexp{$MAPLOOP}, match={XXX}, replace={Fiz. Tekh. Poluprovodn.}]
			\step[fieldsource=\regexp{$MAPLOOP}, match={XXX}, replace={Fiz. Tverd. Tela}]
			\step[fieldsource=\regexp{$MAPLOOP}, match={XXX}, replace={Fiz.-Khim. Mekh. Mater.}]
			\step[fieldsource=\regexp{$MAPLOOP}, match={XXX}, replace={Fizika}]
			\step[fieldsource=\regexp{$MAPLOOP}, match={XXX}, replace={Forsch.-Ber. Landes Nordrh.-Westfal.}]
			\step[fieldsource=\regexp{$MAPLOOP}, match={XXX}, replace={Frequenz}]
			\step[fieldsource=\regexp{$MAPLOOP}, match={XXX}, replace={Fys. Tidsskr.}]
			\step[fieldsource=\regexp{$MAPLOOP}, match={XXX}, replace={G. Fis.}]
			\step[fieldsource=\regexp{$MAPLOOP}, match={XXX}, replace={Geliotekhnika}]
			\step[fieldsource=\regexp{$MAPLOOP}, match={XXX}, replace={Geochim. Cosmochim. Acta}]
			\step[fieldsource=\regexp{$MAPLOOP}, match={XXX}, replace={Haerterei-Tech. Mitt.}]
			\step[fieldsource=\regexp{$MAPLOOP}, match={XXX}, replace={Helv. Chim. Acta}]
			\step[fieldsource=\regexp{$MAPLOOP}, match={XXX}, replace={Helv. Med. Acta}]
			\step[fieldsource=\regexp{$MAPLOOP}, match={XXX}, replace={Helv. Phys. Acta}]
			\step[fieldsource=\regexp{$MAPLOOP}, match={XXX}, replace={Hochfreq. Electroakust}]
			\step[fieldsource=\regexp{$MAPLOOP}, match={XXX}, replace={Hoppe-Seylers Z. Physiol. Chem.}]
			\step[fieldsource=\regexp{$MAPLOOP}, match={XXX}, replace={Inf. Elektron.}]
			\step[fieldsource=\regexp{$MAPLOOP}, match={XXX}, replace={Inf. Elettron.}]
			\step[fieldsource=\regexp{$MAPLOOP}, match={XXX}, replace={Inform. Forsch. Entwickl.}]
			\step[fieldsource=\regexp{$MAPLOOP}, match={XXX}, replace={Inform. Spektrum}]
			\step[fieldsource=\regexp{$MAPLOOP}, match={XXX}, replace={Inform.-Fachber.}]
			\step[fieldsource=\regexp{$MAPLOOP}, match={XXX}, replace={Informatie}]
			\step[fieldsource=\regexp{$MAPLOOP}, match={XXX}, replace={Informatik}]
			\step[fieldsource=\regexp{$MAPLOOP}, match={XXX}, replace={Informatologia Yugosl.}]
			\step[fieldsource=\regexp{$MAPLOOP}, match={XXX}, replace={Informatyka}]
			\step[fieldsource=\regexp{$MAPLOOP}, match={XXX}, replace={Infowelt}]
			\step[fieldsource=\regexp{$MAPLOOP}, match={XXX}, replace={Ing. Electr. Mec.}]
			\step[fieldsource=\regexp{$MAPLOOP}, match={XXX}, replace={Ing. Mec. Electr.}]
			\step[fieldsource=\regexp{$MAPLOOP}, match={XXX}, replace={Ing.-Arch.}]
			\step[fieldsource=\regexp{$MAPLOOP}, match={XXX}, replace={Inzh.-Fiz. Zh.}]
			\step[fieldsource=\regexp{$MAPLOOP}, match={XXX}, replace={Izmer. Tekh.}]
			\step[fieldsource=\regexp{$MAPLOOP}, match={XXX}, replace={Izv. Akad. Nauk Arm. SSR Ser. Tekh. Nauk}]
			\step[fieldsource=\regexp{$MAPLOOP}, match={XXX}, replace={Izv. Akad. Nauk SSSR Energ. Transp.}]
			\step[fieldsource=\regexp{$MAPLOOP}, match={XXX}, replace={Izv. Akad. Nauk SSSR Fiz. Atmos. Okeana}]
			\step[fieldsource=\regexp{$MAPLOOP}, match={XXX}, replace={Izv. Akad. Nauk SSSR Fiz. Zemli}]
			\step[fieldsource=\regexp{$MAPLOOP}, match={XXX}, replace={Izv. Akad. Nauk SSSR Ser. Fiz.}]
			\step[fieldsource=\regexp{$MAPLOOP}, match={XXX}, replace={Izv. Vyssh. Uchebn. Zaved. Elektromekh.}]
			\step[fieldsource=\regexp{$MAPLOOP}, match={XXX}, replace={Izv. Vyssh. Uchebn. Zaved. Radioelektron.}]
			\step[fieldsource=\regexp{$MAPLOOP}, match={XXX}, replace={Izv. Vyssh. Uchebn. Zaved. Radiofiz.}]
			\step[fieldsource=\regexp{$MAPLOOP}, match={XXX}, replace={J. Chim Phys. Phys.-Chim Biol.}]
			\step[fieldsource=\regexp{$MAPLOOP}, match={XXX}, replace={Kernenergie}]
			\step[fieldsource=\regexp{$MAPLOOP}, match={XXX}, replace={Kerntechnik}]
			\step[fieldsource=\regexp{$MAPLOOP}, match={XXX}, replace={Khim. Fiz.}]
			\step[fieldsource=\regexp{$MAPLOOP}, match={XXX}, replace={Kibern. Vychisl. Tekh.}]
			\step[fieldsource=\regexp{$MAPLOOP}, match={XXX}, replace={Kibernetika}]
			\step[fieldsource=\regexp{$MAPLOOP}, match={XXX}, replace={Kristallografiya}]
			\step[fieldsource=\regexp{$MAPLOOP}, match={XXX}, replace={Kvantovaya Elektron. Mosk.}]
			\step[fieldsource=\regexp{$MAPLOOP}, match={XXX}, replace={Kybernetes}]
			\step[fieldsource=\regexp{$MAPLOOP}, match={XXX}, replace={Kybernetika}]
			\step[fieldsource=\regexp{$MAPLOOP}, match={XXX}, replace={Med. Tek.}]
			\step[fieldsource=\regexp{$MAPLOOP}, match={XXX}, replace={Mekh. Avtom. Proizvod.}]
			\step[fieldsource=\regexp{$MAPLOOP}, match={XXX}, replace={Meres Autom.}]
			\step[fieldsource=\regexp{$MAPLOOP}, match={XXX}, replace={Mesures}]
			\step[fieldsource=\regexp{$MAPLOOP}, match={XXX}, replace={Metallofizika}]
			\step[fieldsource=\regexp{$MAPLOOP}, match={XXX}, replace={Metalloved. Term. Obrab. Met.}]
			\step[fieldsource=\regexp{$MAPLOOP}, match={XXX}, replace={Meteorol. Gidrol.}]
			\step[fieldsource=\regexp{$MAPLOOP}, match={XXX}, replace={Meteorol. Rundsch.}]
			\step[fieldsource=\regexp{$MAPLOOP}, match={XXX}, replace={Metrol. Apl.}]
			\step[fieldsource=\regexp{$MAPLOOP}, match={XXX}, replace={Metrologia}]
			\step[fieldsource=\regexp{$MAPLOOP}, match={XXX}, replace={Medel. Simul.}]
			\step[fieldsource=\regexp{$MAPLOOP}, match={XXX}, replace={Nachr. Dok.}]
			\step[fieldsource=\regexp{$MAPLOOP}, match={XXX}, replace={Nachr.tech. Elektron.}]
			\step[fieldsource=\regexp{$MAPLOOP}, match={XXX}, replace={Naturwissentchaften}]
			\step[fieldsource=\regexp{$MAPLOOP}, match={XXX}, replace={Neue Tech.}]
			\step[fieldsource=\regexp{$MAPLOOP}, match={XXX}, replace={Neue Tech. Buero}]
			\step[fieldsource=\regexp{$MAPLOOP}, match={XXX}, replace={Nukleonika}]
			\step[fieldsource=\regexp{$MAPLOOP}, match={XXX}, replace={Numer. Math.}]
			\step[fieldsource=\regexp{$MAPLOOP}, match={XXX}, replace={Nuovo Cimento A}]
			\step[fieldsource=\regexp{$MAPLOOP}, match={XXX}, replace={Nuovo Cimento B}]
			\step[fieldsource=\regexp{$MAPLOOP}, match={XXX}, replace={Nouvo Cimento C}]
			\step[fieldsource=\regexp{$MAPLOOP}, match={XXX}, replace={Nuovo Cimento D}]
			\step[fieldsource=\regexp{$MAPLOOP}, match={XXX}, replace={Okeanologiya}]
			\step[fieldsource=\regexp{$MAPLOOP}, match={XXX}, replace={Opt. Spektrosk.}]
			\step[fieldsource=\regexp{$MAPLOOP}, match={XXX}, replace={Opt.-Mekh. Prom.}]
			\step[fieldsource=\regexp{$MAPLOOP}, match={XXX}, replace={Optik}]
			\step[fieldsource=\regexp{$MAPLOOP}, match={XXX}, replace={Photogrammetria}]
			\step[fieldsource=\regexp{$MAPLOOP}, match={XXX}, replace={Photonics Spectra}]
			\step[fieldsource=\regexp{$MAPLOOP}, match={XXX}, replace={Pis’ma Astron. Zh. }]
			\step[fieldsource=\regexp{$MAPLOOP}, match={XXX}, replace={Pis’ma Zh. Eksp. Teor. Fiz. }]
			\step[fieldsource=\regexp{$MAPLOOP}, match={XXX}, replace={Pis’ma Zh. Tekh. Fiz. }]
			\step[fieldsource=\regexp{$MAPLOOP}, match={XXX}, replace={Poverkhn., Fiz. Khim. Mekh.}]
			\step[fieldsource=\regexp{$MAPLOOP}, match={XXX}, replace={Pr. Inst. Elektrotech.}]
			\step[fieldsource=\regexp{$MAPLOOP}, match={XXX}, replace={Prib. Sist. Upr.}]
			\step[fieldsource=\regexp{$MAPLOOP}, match={XXX}, replace={Prib. Tekh. Eksp.}]
			\step[fieldsource=\regexp{$MAPLOOP}, match={XXX}, replace={Prikl. Mat. Mekh.}]
			\step[fieldsource=\regexp{$MAPLOOP}, match={XXX}, replace={Prikl. Mekh.}]
			\step[fieldsource=\regexp{$MAPLOOP}, match={XXX}, replace={Probl. Kibern.}]
			\step[fieldsource=\regexp{$MAPLOOP}, match={XXX}, replace={Probl. Peredachi Inf.}]
			\step[fieldsource=\regexp{$MAPLOOP}, match={XXX}, replace={Proc. K. Ned. Akad. Wet. B, Palaeontol.}]
			\step[fieldsource=\regexp{$MAPLOOP}, match={XXX}, replace={Anthropol. }]
			\step[fieldsource=\regexp{$MAPLOOP}, match={XXX}, replace={Programmirovanie}]
			\step[fieldsource=\regexp{$MAPLOOP}, match={XXX}, replace={Prz. Elektrotech.}]
			\step[fieldsource=\regexp{$MAPLOOP}, match={XXX}, replace={Prz. Telekomun.}]
			\step[fieldsource=\regexp{$MAPLOOP}, match={XXX}, replace={PT/Elektrotech. Elektron.}]
			\step[fieldsource=\regexp{$MAPLOOP}, match={XXX}, replace={Radio Fernsehen Elektron.}]
			\step[fieldsource=\regexp{$MAPLOOP}, match={XXX}, replace={Radiotekh. Elektron.}]
			\step[fieldsource=\regexp{$MAPLOOP}, match={XXX}, replace={Radiotekhnika Mosk.}]
			\step[fieldsource=\regexp{$MAPLOOP}, match={XXX}, replace={Rev. Acad. Cienc. Zaragoza}]
			\step[fieldsource=\regexp{$MAPLOOP}, match={XXX}, replace={Rev. Electrotec. (Argentina)}]
			\step[fieldsource=\regexp{$MAPLOOP}, match={XXX}, replace={Rev. Electrotec. (Spain)}]
			\step[fieldsource=\regexp{$MAPLOOP}, match={XXX}, replace={Rev. Energ.}]
			\step[fieldsource=\regexp{$MAPLOOP}, match={XXX}, replace={Rev. Esp. Electron. Rev. Geofis.}]
			\step[fieldsource=\regexp{$MAPLOOP}, match={XXX}, replace={Ric. Autom.}]
			\step[fieldsource=\regexp{$MAPLOOP}, match={XXX}, replace={Ric. Spettrosc.}]
			\step[fieldsource=\regexp{$MAPLOOP}, match={XXX}, replace={Robotersysteme}]
			\step[fieldsource=\regexp{$MAPLOOP}, match={XXX}, replace={Rozpr. Electrotech.}]
			\step[fieldsource=\regexp{$MAPLOOP}, match={XXX}, replace={Sadhana}]
			\step[fieldsource=\regexp{$MAPLOOP}, match={XXX}, replace={Schweiz. Tech. Z.}]
			\step[fieldsource=\regexp{$MAPLOOP}, match={XXX}, replace={Scientia}]
			\step[fieldsource=\regexp{$MAPLOOP}, match={XXX}, replace={Siemens Forsch. Entwickl. Ber.}]
			\step[fieldsource=\regexp{$MAPLOOP}, match={XXX}, replace={Sist. Autom.}]
			\step[fieldsource=\regexp{$MAPLOOP}, match={XXX}, replace={Sitzungsber. Oester. Akad. Wiss. Math.-}]
			\step[fieldsource=\regexp{$MAPLOOP}, match={XXX}, replace={Naturwiss. Kl. Abt. II (Austria)}]
			\step[fieldsource=\regexp{$MAPLOOP}, match={XXX}, replace={Spectrochim. Acta A, Mol. Spectrosc.}]
			\step[fieldsource=\regexp{$MAPLOOP}, match={XXX}, replace={Spectrochim. Acta B, At. Spectrosc.}]
			\step[fieldsource=\regexp{$MAPLOOP}, match={XXX}, replace={Sprache Datenverarb.}]
			\step[fieldsource=\regexp{$MAPLOOP}, match={XXX}, replace={Stanki Instrum.}]
			\step[fieldsource=\regexp{$MAPLOOP}, match={XXX}, replace={Steklo Keram.}]
			\step[fieldsource=\regexp{$MAPLOOP}, match={XXX}, replace={Svetotekhnika  TE Int.}]
			\step[fieldsource=\regexp{$MAPLOOP}, match={XXX}, replace={Tech. Bull. Vevey}]
			\step[fieldsource=\regexp{$MAPLOOP}, match={XXX}, replace={Tech. Mitt. Krupp (Engl. Ed.)}]
			\step[fieldsource=\regexp{$MAPLOOP}, match={XXX}, replace={Tech. Mitt. PTT}]
			\step[fieldsource=\regexp{$MAPLOOP}, match={XXX}, replace={Tech. Mitt. RFZ}]
			\step[fieldsource=\regexp{$MAPLOOP}, match={XXX}, replace={Technica}]
			\step[fieldsource=\regexp{$MAPLOOP}, match={XXX}, replace={Tecnica}]
			\step[fieldsource=\regexp{$MAPLOOP}, match={XXX}, replace={Teh. Fiz.}]
			\step[fieldsource=\regexp{$MAPLOOP}, match={XXX}, replace={Tehnika}]
			\step[fieldsource=\regexp{$MAPLOOP}, match={XXX}, replace={Tekh. Elektrodin.}]
			\step[fieldsource=\regexp{$MAPLOOP}, match={XXX}, replace={Tekh. Kibern.}]
			\step[fieldsource=\regexp{$MAPLOOP}, match={XXX}, replace={Tekh. Kino Telev.}]
			\step[fieldsource=\regexp{$MAPLOOP}, match={XXX}, replace={Tekh. Misul}]
			\step[fieldsource=\regexp{$MAPLOOP}, match={XXX}, replace={Telekomunikacije}]
			\step[fieldsource=\regexp{$MAPLOOP}, match={XXX}, replace={Telektronikk}]
			\step[fieldsource=\regexp{$MAPLOOP}, match={XXX}, replace={Teleteknik}]
			\step[fieldsource=\regexp{$MAPLOOP}, match={XXX}, replace={Teor. Mat. Fiz.}]
			\step[fieldsource=\regexp{$MAPLOOP}, match={XXX}, replace={Teploeoergetika}]
			\step[fieldsource=\regexp{$MAPLOOP}, match={XXX}, replace={Teplofiz. Vvs. Temp.}]
			\step[fieldsource=\regexp{$MAPLOOP}, match={XXX}, replace={Tidskr. Dok.}]
			\step[fieldsource=\regexp{$MAPLOOP}, match={XXX}, replace={TN Nachr.}]
			\step[fieldsource=\regexp{$MAPLOOP}, match={XXX}, replace={Toute Electron.}]
			\step[fieldsource=\regexp{$MAPLOOP}, match={XXX}, replace={Tr. Inst. Teor. Astron.}]
			\step[fieldsource=\regexp{$MAPLOOP}, match={XXX}, replace={Ukr. Fiz. Zh.}]
			\step[fieldsource=\regexp{$MAPLOOP}, match={XXX}, replace={Usp. Fiz. Nauk}]
			\step[fieldsource=\regexp{$MAPLOOP}, match={XXX}, replace={Vak.-Tech.}]
			\step[fieldsource=\regexp{$MAPLOOP}, match={XXX}, replace={VDE Fachiber.}]
			\step[fieldsource=\regexp{$MAPLOOP}, match={XXX}, replace={VDI Z.}]
			\step[fieldsource=\regexp{$MAPLOOP}, match={XXX}, replace={Vestn. Mashinostr.}]
			\step[fieldsource=\regexp{$MAPLOOP}, match={XXX}, replace={Vestn. Mosk. Univ. 15, Vychisl. Mat. Kibern.}]
			\step[fieldsource=\regexp{$MAPLOOP}, match={XXX}, replace={Vestn. Mosk. Univ. 3, Fiz. Astron.}]
			\step[fieldsource=\regexp{$MAPLOOP}, match={XXX}, replace={Vesti Akad. Navuk BSSR Ser. Fiz. Energ. Navuk}]
			\step[fieldsource=\regexp{$MAPLOOP}, match={XXX}, replace={VGB Kraftwerkstech. (Ger. Ed.)}]
			\step[fieldsource=\regexp{$MAPLOOP}, match={XXX}, replace={Vide Couches Minces}]
			\step[fieldsource=\regexp{$MAPLOOP}, match={XXX}, replace={Vistas Astron.}]
			\step[fieldsource=\regexp{$MAPLOOP}, match={XXX}, replace={Vopr. At. Nauki Tekh. Ser., Fiz. Radiats.}]
			\step[fieldsource=\regexp{$MAPLOOP}, match={XXX}, replace={Povrezhdenii Radiats. Materialoved.}]
			\step[fieldsource=\regexp{$MAPLOOP}, match={XXX}, replace={Vopr. At. Nauki Tekh. Ser., Obshch. Yad. Fiz.}]
			\step[fieldsource=\regexp{$MAPLOOP}, match={XXX}, replace={Vuoto Sci. Tecnol.}]
			\step[fieldsource=\regexp{$MAPLOOP}, match={XXX}, replace={Wiss. Z. Friedrich-Schiller-Univ. Jena}]
			\step[fieldsource=\regexp{$MAPLOOP}, match={XXX}, replace={Nat.wiss. Reihe}]
			\step[fieldsource=\regexp{$MAPLOOP}, match={XXX}, replace={Wiss. Z. Karl-Marx-Univ. Leipz. Math.-}]
			\step[fieldsource=\regexp{$MAPLOOP}, match={XXX}, replace={Nat.wiss. Reihe}]
			\step[fieldsource=\regexp{$MAPLOOP}, match={XXX}, replace={Wiss. Z. Tech. Hochsch. Ilmenau}]
			\step[fieldsource=\regexp{$MAPLOOP}, match={XXX}, replace={Wiss. Z. Tech. Univ. Dresd.}]
			\step[fieldsource=\regexp{$MAPLOOP}, match={XXX}, replace={Wiss. Z. Tech. Univ. Karl-Marx-Stadt}]
			\step[fieldsource=\regexp{$MAPLOOP}, match={XXX}, replace={Yad. Fiz.}]
			\step[fieldsource=\regexp{$MAPLOOP}, match={XXX}, replace={Z. Angew. Math. Mech.}]
			\step[fieldsource=\regexp{$MAPLOOP}, match={XXX}, replace={Z. Angew. Math. Phys.}]
			\step[fieldsource=\regexp{$MAPLOOP}, match={XXX}, replace={Z. Met.kd.}]
			\step[fieldsource=\regexp{$MAPLOOP}, match={XXX}, replace={Z. Nat. Forsch. A, Phys. Phys. Chem. Kosmophys.}]
			\step[fieldsource=\regexp{$MAPLOOP}, match={XXX}, replace={Z. Oper. Res. A, Theor.}]
			\step[fieldsource=\regexp{$MAPLOOP}, match={XXX}, replace={Z. Oper. Res. B, Prax.}]
			\step[fieldsource=\regexp{$MAPLOOP}, match={XXX}, replace={Z. Phys.}]
			\step[fieldsource=\regexp{$MAPLOOP}, match={XXX}, replace={Z. Phys. A, At. Nuclei}]
			\step[fieldsource=\regexp{$MAPLOOP}, match={XXX}, replace={Z. Phys. B, Condens. Matter}]
			\step[fieldsource=\regexp{$MAPLOOP}, match={XXX}, replace={Z. Phys. C, Part Fields}]
			\step[fieldsource=\regexp{$MAPLOOP}, match={XXX}, replace={Z. Phys. Chem. Neue Folge}]
			\step[fieldsource=\regexp{$MAPLOOP}, match={XXX}, replace={Z. Phys. Chem., Leipz.}]
			\step[fieldsource=\regexp{$MAPLOOP}, match={XXX}, replace={Z. Phys. D, At. Mol. Clusters}]
			\step[fieldsource=\regexp{$MAPLOOP}, match={XXX}, replace={Zavod. Lab.}]
			\step[fieldsource=\regexp{$MAPLOOP}, match={XXX}, replace={Zh. Eksp. Teor. Fiz.}]
			\step[fieldsource=\regexp{$MAPLOOP}, match={XXX}, replace={Zh. Fiz. Khim.}]
			\step[fieldsource=\regexp{$MAPLOOP}, match={XXX}, replace={Zh. Prikl. Mekh. Tekh. Fiz.}]
			\step[fieldsource=\regexp{$MAPLOOP}, match={XXX}, replace={Zh. Prikl. Spektrosk.}]
			\step[fieldsource=\regexp{$MAPLOOP}, match={XXX}, replace={Zh. Tekh. Fiz.}]
			\step[fieldsource=\regexp{$MAPLOOP}, match={XXX}, replace={Zh. Vychisl. Mat. Mat. Fiz.}]
			\step[fieldsource=\regexp{$MAPLOOP}, match={XXX}, replace={Zisin, J. Seismol. Soc. Jpn.}] 
			\step[fieldsource=\regexp{$MAPLOOP}, match={Production}, replace={Prod.}]
			\step[fieldsource=\regexp{$MAPLOOP}, match={Reliability}, replace={Rel.}]
			\step[fieldsource=\regexp{$MAPLOOP}, match={Report}, replace={Rep.}]
			\step[fieldsource=\regexp{$MAPLOOP}, match={Semiconductor}, replace={Semicond.}]
			\step[fieldsource=\regexp{$MAPLOOP}, match={Research}, replace={Res.}]
			\step[fieldsource=\regexp{$MAPLOOP}, match={Sensing}, replace={Sens.}]
			\step[fieldsource=\regexp{$MAPLOOP}, match={Resonance}, replace={Reson.}]
			\step[fieldsource=\regexp{$MAPLOOP}, match={Series}, replace={Ser.}]
			\step[fieldsource=\regexp{$MAPLOOP}, match={Resources}, replace={Resour.}]
			\step[fieldsource=\regexp{$MAPLOOP}, match={Simulation}, replace={Simul.}]
			\step[fieldsource=\regexp{$MAPLOOP}, match={Reviews}, replace={Rev.}]
			\step[fieldsource=\regexp{$MAPLOOP}, match={Review}, replace={Rev.}]
			\step[fieldsource=\regexp{$MAPLOOP}, match={Singapore}, replace={Singap.}]
			\step[fieldsource=\regexp{$MAPLOOP}, match={Robotics}, replace={Robot.}]
			\step[fieldsource=\regexp{$MAPLOOP}, match={Sistema}, replace={Sist.}]
			\step[fieldsource=\regexp{$MAPLOOP}, match={Royal}, replace={Roy.}]
			\step[fieldsource=\regexp{$MAPLOOP}, match={Society}, replace={Soc.}]
			\step[fieldsource=\regexp{$MAPLOOP}, match={Safety}, replace={Saf.}]
			\step[fieldsource=\regexp{$MAPLOOP}, match={Sociological}, replace={Sociol.}]
			\step[fieldsource=\regexp{$MAPLOOP}, match={Satellite}, replace={Satell.}]
			\step[fieldsource=\regexp{$MAPLOOP}, match={Software}, replace={Softw.}]
			\step[fieldsource=\regexp{$MAPLOOP}, match={Scandinavian}, replace={Scand.}]
			\step[fieldsource=\regexp{$MAPLOOP}, match={Solar}, replace={Sol.}]
			\step[fieldsource=\regexp{$MAPLOOP}, match={Sciences}, replace={Sci.}]
			\step[fieldsource=\regexp{$MAPLOOP}, match={Science}, replace={Sci.}]
			\step[fieldsource=\regexp{$MAPLOOP}, match={Soviet}, replace={Sov.}]
			\step[fieldsource=\regexp{$MAPLOOP}, match={Section}, replace={Sect.}]
			\step[fieldsource=\regexp{$MAPLOOP}, match={Spectroscopy}, replace={Spectrosc.}]
			\step[fieldsource=\regexp{$MAPLOOP}, match={Security}, replace={Secur.}]
			\step[fieldsource=\regexp{$MAPLOOP}, match={Spectrum}, replace={Spectr.}]
			\step[fieldsource=\regexp{$MAPLOOP}, match={Seismology}, replace={Seismol.}]
			\step[fieldsource=\regexp{$MAPLOOP}, match={Speculations}, replace={Specul.}]
			\step[fieldsource=\regexp{$MAPLOOP}, match={Selected}, replace={Sel.}]
			\step[fieldsource=\regexp{$MAPLOOP}, match={Statistics}, replace={Statist.}]
			\step[fieldsource=\regexp{$MAPLOOP}, match={Structures}, replace={Struct.}]
			\step[fieldsource=\regexp{$MAPLOOP}, match={Structure}, replace={Struct.}]
			\step[fieldsource=\regexp{$MAPLOOP}, match={Terrestrial}, replace={Terr.}]
			\step[fieldsource=\regexp{$MAPLOOP}, match={Studies}, replace={Stud.}]
			\step[fieldsource=\regexp{$MAPLOOP}, match={Theoretical}, replace={Theor.}]
			\step[fieldsource=\regexp{$MAPLOOP}, match={Superconductivity}, replace={Supercond.}]
			\step[fieldsource=\regexp{$MAPLOOP}, match={Transactions}, replace={Trans.}]
			\step[fieldsource=\regexp{$MAPLOOP}, match={Supplement}, replace={Suppl.}]
			\step[fieldsource=\regexp{$MAPLOOP}, match={Translation}, replace={Transl.}]
			\step[fieldsource=\regexp{$MAPLOOP}, match={Surface}, replace={Surf.}]
			\step[fieldsource=\regexp{$MAPLOOP}, match={Transmission}, replace={Transmiss.}]
			\step[fieldsource=\regexp{$MAPLOOP}, match={Survey}, replace={Surv.}]
			\step[fieldsource=\regexp{$MAPLOOP}, match={Transportation}, replace={Transp.}]
			\step[fieldsource=\regexp{$MAPLOOP}, match={Sustainable}, replace={Sustain.}]
			\step[fieldsource=\regexp{$MAPLOOP}, match={Tutorials}, replace={Tut.}]
			\step[fieldsource=\regexp{$MAPLOOP}, match={Symposium}, replace={Symp.}]
			\step[fieldsource=\regexp{$MAPLOOP}, match={Ultrasonic}, replace={Ultrason.}]
			\step[fieldsource=\regexp{$MAPLOOP}, match={Systems}, replace={Syst.}]
			\step[fieldsource=\regexp{$MAPLOOP}, match={System}, replace={Syst.}]
			\step[fieldsource=\regexp{$MAPLOOP}, match={University}, replace={Univ.}]
			\step[fieldsource=\regexp{$MAPLOOP}, match={\detokenize{Universität}}, replace={Univ.}]
			\step[fieldsource=\regexp{$MAPLOOP}, match={\detokenize{Université}}, replace={Univ.}]
			\step[fieldsource=\regexp{$MAPLOOP}, match={Vacuum}, replace={Vac.}]
			\step[fieldsource=\regexp{$MAPLOOP}, match={Vehicles}, replace={Veh.}]
			\step[fieldsource=\regexp{$MAPLOOP}, match={Vehicle}, replace={Veh.}]
			\step[fieldsource=\regexp{$MAPLOOP}, match={Vehicular}, replace={Veh.}]
			\step[fieldsource=\regexp{$MAPLOOP}, match={Technology}, replace={Technol.}]
			\step[fieldsource=\regexp{$MAPLOOP}, match={Technological}, replace={Technol.}]
			\step[fieldsource=\regexp{$MAPLOOP}, match={Vibration}, replace={Vib.}]
			\step[fieldsource=\regexp{$MAPLOOP}, match={Telecommunications}, replace={Telecommun.}]
			\step[fieldsource=\regexp{$MAPLOOP}, match={Visual}, replace={Vis.}]
			\step[fieldsource=\regexp{$MAPLOOP}, match={Television}, replace={Telev.}]
			\step[fieldsource=\regexp{$MAPLOOP}, match={Welding}, replace={Weld.}]
			\step[fieldsource=\regexp{$MAPLOOP}, match={Temperature}, replace={Temp.}]
			\step[fieldsource=\regexp{$MAPLOOP}, match={Working}, replace={Work.}]
			\step[fieldsource=\regexp{$MAPLOOP}, match={Learning}, replace={Learn.}]
			\step[fieldsource=\regexp{$MAPLOOP}, match={Measurement}, replace={Meas.}]
			\step[fieldsource=\regexp{$MAPLOOP}, match={Letters}, replace={Lett.}]
			\step[fieldsource=\regexp{$MAPLOOP}, match={Letter}, replace={Lett.}]
			\step[fieldsource=\regexp{$MAPLOOP}, match={Mechanical}, replace={Mech.}]
			\step[fieldsource=\regexp{$MAPLOOP}, match={Mechanics}, replace={Mech.}]
			\step[fieldsource=\regexp{$MAPLOOP}, match={Mechanic}, replace={Mech.}]
			\step[fieldsource=\regexp{$MAPLOOP}, match={Lightwave}, replace={Lightw.}]
			\step[fieldsource=\regexp{$MAPLOOP}, match={Medical}, replace={Med.}]
			\step[fieldsource=\regexp{$MAPLOOP}, match={Logic, Logical}, replace={Log.}]
			\step[fieldsource=\regexp{$MAPLOOP}, match={Metals}, replace={Met.}]
			\step[fieldsource=\regexp{$MAPLOOP}, match={Luminescence}, replace={Lumin.}]
			\step[fieldsource=\regexp{$MAPLOOP}, match={Metallurgy}, replace={Metall.}]
			\step[fieldsource=\regexp{$MAPLOOP}, match={Machines}, replace={Mach.}]
			\step[fieldsource=\regexp{$MAPLOOP}, match={Machine}, replace={Mach.}]
			\step[fieldsource=\regexp{$MAPLOOP}, match={Meteorology}, replace={Meteorol.}]
			\step[fieldsource=\regexp{$MAPLOOP}, match={Magazine}, replace={Mag.}]
			\step[fieldsource=\regexp{$MAPLOOP}, match={Metropolitan}, replace={Metrop.}]
			\step[fieldsource=\regexp{$MAPLOOP}, match={Magnetics}, replace={Magn.}]
			\step[fieldsource=\regexp{$MAPLOOP}, match={Mexican, Mexico}, replace={Mex.}]
			\step[fieldsource=\regexp{$MAPLOOP}, match={Management}, replace={Manage.}]
			\step[fieldsource=\regexp{$MAPLOOP}, match={Microelectromechanical}, replace={Microelectromech.}]
			\step[fieldsource=\regexp{$MAPLOOP}, match={Managing}, replace={Manag.}]
			\step[fieldsource=\regexp{$MAPLOOP}, match={Microgravity}, replace={Microgr.}]
			\step[fieldsource=\regexp{$MAPLOOP}, match={Manufacturing}, replace={Manuf.}]
			\step[fieldsource=\regexp{$MAPLOOP}, match={Microscopy}, replace={Microsc.}]
			\step[fieldsource=\regexp{$MAPLOOP}, match={Marine}, replace={Mar.}]
			\step[fieldsource=\regexp{$MAPLOOP}, match={Microwaves}, replace={Microw.}]
			\step[fieldsource=\regexp{$MAPLOOP}, match={Microwave}, replace={Microw.}]
			\step[fieldsource=\regexp{$MAPLOOP}, match={Materials}, replace={Mater.}]
			\step[fieldsource=\regexp{$MAPLOOP}, match={Material}, replace={Mater.}]
			\step[fieldsource=\regexp{$MAPLOOP}, match={Military}, replace={Mil.}]
			\step[fieldsource=\regexp{$MAPLOOP}, match={Modeling}, replace={Model.}]
			\step[fieldsource=\regexp{$MAPLOOP}, match={Modelling}, replace={Model.}]
			\step[fieldsource=\regexp{$MAPLOOP}, match={Oceanic}, replace={Ocean.}]
			\step[fieldsource=\regexp{$MAPLOOP}, match={Molecular}, replace={Mol.}]
			\step[fieldsource=\regexp{$MAPLOOP}, match={Oceanography}, replace={Oceanogr.}]
			\step[fieldsource=\regexp{$MAPLOOP}, match={Monitoring}, replace={Monit.}]
			\step[fieldsource=\regexp{$MAPLOOP}, match={Occupation}, replace={Occupat.}]
			\step[fieldsource=\regexp{$MAPLOOP}, match={Multiphysics}, replace={Multiphys.}]
			\step[fieldsource=\regexp{$MAPLOOP}, match={Operational}, replace={Oper.}]
			\step[fieldsource=\regexp{$MAPLOOP}, match={Nanobioscience}, replace={Nanobiosci.}]
			\step[fieldsource=\regexp{$MAPLOOP}, match={Optical}, replace={Opt.}]
			\step[fieldsource=\regexp{$MAPLOOP}, match={Nanotechnology}, replace={Nanotechnol.}]
			\step[fieldsource=\regexp{$MAPLOOP}, match={Optics}, replace={Opt.}]
			\step[fieldsource=\regexp{$MAPLOOP}, match={National}, replace={Nat.}]
			\step[fieldsource=\regexp{$MAPLOOP}, match={Optimization}, replace={Optim.}]
			\step[fieldsource=\regexp{$MAPLOOP}, match={Naval}, replace={Nav.}]
			\step[fieldsource=\regexp{$MAPLOOP}, match={Organization}, replace={Org.}]
			\step[fieldsource=\regexp{$MAPLOOP}, match={Networking}, replace={Netw.}]
			\step[fieldsource=\regexp{$MAPLOOP}, match={Networked}, replace={Netw.}]
			\step[fieldsource=\regexp{$MAPLOOP}, match={Network}, replace={Netw.}]
			\step[fieldsource=\regexp{$MAPLOOP}, match={Packaging}, replace={Packag.}]
			\step[fieldsource=\regexp{$MAPLOOP}, match={Newsletter}, replace={Newslett.}]
			\step[fieldsource=\regexp{$MAPLOOP}, match={Particle}, replace={Part.}]
			\step[fieldsource=\regexp{$MAPLOOP}, match={Nondestructive}, replace={Nondestruct.}]
			\step[fieldsource=\regexp{$MAPLOOP}, match={Patent}, replace={Pat.}]
			\step[fieldsource=\regexp{$MAPLOOP}, match={Nuclear}, replace={Nucl.}]
			\step[fieldsource=\regexp{$MAPLOOP}, match={Performance}, replace={Perform.}]
			\step[fieldsource=\regexp{$MAPLOOP}, match={Numerical}, replace={Numer.}]
			\step[fieldsource=\regexp{$MAPLOOP}, match={Personal}, replace={Pers.}]
			\step[fieldsource=\regexp{$MAPLOOP}, match={Observations}, replace={Observ.}]
			\step[fieldsource=\regexp{$MAPLOOP}, match={Philosophical}, replace={Philos.}]
			\step[fieldsource=\regexp{$MAPLOOP}, match={Photonics}, replace={Photon.}]
			\step[fieldsource=\regexp{$MAPLOOP}, match={Productivity}, replace={Productiv.}]
			\step[fieldsource=\regexp{$MAPLOOP}, match={Photovoltaics}, replace={Photovolt.}]
			\step[fieldsource=\regexp{$MAPLOOP}, match={Programming}, replace={Program.}]
			\step[fieldsource=\regexp{$MAPLOOP}, match={Physics}, replace={Phys.}]
			\step[fieldsource=\regexp{$MAPLOOP}, match={Progress}, replace={Prog.}]
			\step[fieldsource=\regexp{$MAPLOOP}, match={Physiology}, replace={Physiol.}]
			\step[fieldsource=\regexp{$MAPLOOP}, match={Propagation}, replace={Propag.}]
			\step[fieldsource=\regexp{$MAPLOOP}, match={Planetary}, replace={Planet.}]
			\step[fieldsource=\regexp{$MAPLOOP}, match={Psychology}, replace={Psychol.}]
			\step[fieldsource=\regexp{$MAPLOOP}, match={Pneumatics}, replace={Pneum.}]
			\step[fieldsource=\regexp{$MAPLOOP}, match={Quality}, replace={Qual.}]
			\step[fieldsource=\regexp{$MAPLOOP}, match={Pollution}, replace={Pollut.}]
			\step[fieldsource=\regexp{$MAPLOOP}, match={Quarterly}, replace={Quart.}]
			\step[fieldsource=\regexp{$MAPLOOP}, match={Polymer}, replace={Polym.}]
			\step[fieldsource=\regexp{$MAPLOOP}, match={Radiation}, replace={Radiat.}]
			\step[fieldsource=\regexp{$MAPLOOP}, match={Polytechnic}, replace={Polytech.}]
			\step[fieldsource=\regexp{$MAPLOOP}, match={Radiology}, replace={Radiol.}]
			\step[fieldsource=\regexp{$MAPLOOP}, match={Practice}, replace={Pract.}]
			\step[fieldsource=\regexp{$MAPLOOP}, match={Reactor}, replace={React.}]
			\step[fieldsource=\regexp{$MAPLOOP}, match={Precision}, replace={Precis.}]
			\step[fieldsource=\regexp{$MAPLOOP}, match={Receivers}, replace={Receiv.}]
			\step[fieldsource=\regexp{$MAPLOOP}, match={Principles}, replace={Princ.}]
			\step[fieldsource=\regexp{$MAPLOOP}, match={Recognition}, replace={Recognit.}]
			\step[fieldsource=\regexp{$MAPLOOP}, match={Proceedings}, replace={Proc.}]
			\step[fieldsource=\regexp{$MAPLOOP}, match={Record}, replace={Rec.}]
			\step[fieldsource=\regexp{$MAPLOOP}, match={Processing}, replace={Process.}]
			\step[fieldsource=\regexp{$MAPLOOP}, match={Rehabilitation}, replace={Rehabil.}]
			\step[fieldsource=\regexp{$MAPLOOP}, match={Conversion}, replace={Convers.}]
			\step[fieldsource=\regexp{$MAPLOOP}, match={Digital}, replace={Digit.}]
			\step[fieldsource=\regexp{$MAPLOOP}, match={Convention}, replace={Conv.}]
			\step[fieldsource=\regexp{$MAPLOOP}, match={Disclosure}, replace={Discl.}]
			\step[fieldsource=\regexp{$MAPLOOP}, match={Correspondence}, replace={Corresp.}]
			\step[fieldsource=\regexp{$MAPLOOP}, match={Discussions}, replace={Discuss.}]
			\step[fieldsource=\regexp{$MAPLOOP}, match={Critical}, replace={Crit.}]
			\step[fieldsource=\regexp{$MAPLOOP}, match={Dissertations}, replace={Diss.}]
			\step[fieldsource=\regexp{$MAPLOOP}, match={Crystal}, replace={Cryst.}]
			\step[fieldsource=\regexp{$MAPLOOP}, match={Distributed}, replace={Distrib.}]
			\step[fieldsource=\regexp{$MAPLOOP}, match={Crystallography}, replace={Crystallogr.}]
			\step[fieldsource=\regexp{$MAPLOOP}, match={Dynamics}, replace={Dyn.}]
			\step[fieldsource=\regexp{$MAPLOOP}, match={Cybernetics}, replace={Cybern.}]
			\step[fieldsource=\regexp{$MAPLOOP}, match={Earthquake}, replace={Earthq.}]
			\step[fieldsource=\regexp{$MAPLOOP}, match={Decision}, replace={Decis.}]
			\step[fieldsource=\regexp{$MAPLOOP}, match={Economics}, replace={Econ.}]
			\step[fieldsource=\regexp{$MAPLOOP}, match={Economic}, replace={Econ.}]
			\step[fieldsource=\regexp{$MAPLOOP}, match={Economical}, replace={Econ.}]
			\step[fieldsource=\regexp{$MAPLOOP}, match={Edition}, replace={Ed.}]
			\step[fieldsource=\regexp{$MAPLOOP}, match={Evolutionary}, replace={Evol.}]
			\step[fieldsource=\regexp{$MAPLOOP}, match={Education}, replace={Educ.}]
			\step[fieldsource=\regexp{$MAPLOOP}, match={Exhibition}, replace={Exhib.}]
			\step[fieldsource=\regexp{$MAPLOOP}, match={Electrical}, replace={Elect.}]
			\step[fieldsource=\regexp{$MAPLOOP}, match={Electric}, replace={Elect.}]
			\step[fieldsource=\regexp{$MAPLOOP}, match={Experimental}, replace={Exp.}]
			\step[fieldsource=\regexp{$MAPLOOP}, match={Electrification}, replace={Electrific.}]
			\step[fieldsource=\regexp{$MAPLOOP}, match={Exploratory}, replace={Explor.}]
			\step[fieldsource=\regexp{$MAPLOOP}, match={Electromagnetic}, replace={Electromagn.}]
			\step[fieldsource=\regexp{$MAPLOOP}, match={Exposition}, replace={Expo.}]
			\step[fieldsource=\regexp{$MAPLOOP}, match={Electroacoustic}, replace={Electroacoust.}]
			\step[fieldsource=\regexp{$MAPLOOP}, match={Express}, replace={Express}]
			\step[fieldsource=\regexp{$MAPLOOP}, match={Electronics}, replace={Electron.}]
			\step[fieldsource=\regexp{$MAPLOOP}, match={Electronic}, replace={Electron.}]
			\step[fieldsource=\regexp{$MAPLOOP}, match={Fabrication}, replace={Fabr.}]
			\step[fieldsource=\regexp{$MAPLOOP}, match={Emerging}, replace={Emerg.}]
			\step[fieldsource=\regexp{$MAPLOOP}, match={Faculty}, replace={Fac.}]
			\step[fieldsource=\regexp{$MAPLOOP}, match={Engineering}, replace={Eng.}]
			\step[fieldsource=\regexp{$MAPLOOP}, match={Engineers}, replace={Eng.}]
			\step[fieldsource=\regexp{$MAPLOOP}, match={Engineer}, replace={Eng.}]
			\step[fieldsource=\regexp{$MAPLOOP}, match={Ferroelectrics}, replace={Ferroelect.}]
			\step[fieldsource=\regexp{$MAPLOOP}, match={Environment}, replace={Environ.}]
			\step[fieldsource=\regexp{$MAPLOOP}, match={Francais, French}, replace={Fr.}]
			\step[fieldsource=\regexp{$MAPLOOP}, match={Equations}, replace={Equ.}]
			\step[fieldsource=\regexp{$MAPLOOP}, match={Frequency}, replace={Freq.}]
			\step[fieldsource=\regexp{$MAPLOOP}, match={Equipment}, replace={Equip.}]
			\step[fieldsource=\regexp{$MAPLOOP}, match={Foundation}, replace={Found.}]
			\step[fieldsource=\regexp{$MAPLOOP}, match={Ergonomics}, replace={Ergonom.}]
			\step[fieldsource=\regexp{$MAPLOOP}, match={Fundamental}, replace={Fundam.}]
			\step[fieldsource=\regexp{$MAPLOOP}, match={European}, replace={Eur.}]
			\step[fieldsource=\regexp{$MAPLOOP}, match={Generation}, replace={Gener.}]
			\step[fieldsource=\regexp{$MAPLOOP}, match={Evaluation}, replace={Eval.}]
			\step[fieldsource=\regexp{$MAPLOOP}, match={Geology}, replace={Geol.}]
			\step[fieldsource=\regexp{$MAPLOOP}, match={Geophysics}, replace={Geophys.}]
			\step[fieldsource=\regexp{$MAPLOOP}, match={Innovations}, replace={Innov.}]
			\step[fieldsource=\regexp{$MAPLOOP}, match={Innovation}, replace={Innov.}]
			\step[fieldsource=\regexp{$MAPLOOP}, match={Geoscience}, replace={Geosci.}]
			\step[fieldsource=\regexp{$MAPLOOP}, match={Institute}, replace={Inst.}]
			\step[fieldsource=\regexp{$MAPLOOP}, match={Institution}, replace={Inst.}]
			\step[fieldsource=\regexp{$MAPLOOP}, match={Graphics}, replace={Graph.}]
			\step[fieldsource=\regexp{$MAPLOOP}, match={Instrument}, replace={Instrum.}]
			\step[fieldsource=\regexp{$MAPLOOP}, match={Guidance}, replace={Guid.}]
			\step[fieldsource=\regexp{$MAPLOOP}, match={Instrumentation}, replace={Instrum.}]
			\step[fieldsource=\regexp{$MAPLOOP}, match={Harmonics}, replace={Harmon.}]
			\step[fieldsource=\regexp{$MAPLOOP}, match={Harmonic}, replace={Harmon.}]
			\step[fieldsource=\regexp{$MAPLOOP}, match={Insulation}, replace={Insul.}]
			\step[fieldsource=\regexp{$MAPLOOP}, match={History}, replace={Hist.}]
			\step[fieldsource=\regexp{$MAPLOOP}, match={Integrated}, replace={Integr.}]
			\step[fieldsource=\regexp{$MAPLOOP}, match={Horizon}, replace={Horiz.}]
			\step[fieldsource=\regexp{$MAPLOOP}, match={Intelligence}, replace={Intell.}]
			\step[fieldsource=\regexp{$MAPLOOP}, match={Hungary}, replace={Hung.}]
			\step[fieldsource=\regexp{$MAPLOOP}, match={Hungarian}, replace={Hung.}]
			\step[fieldsource=\regexp{$MAPLOOP}, match={Intelligent}, replace={Intell.}]
			\step[fieldsource=\regexp{$MAPLOOP}, match={Hydraulics}, replace={Hydraul.}]
			\step[fieldsource=\regexp{$MAPLOOP}, match={Interactions}, replace={Interact.}]
			\step[fieldsource=\regexp{$MAPLOOP}, match={Hydrology}, replace={Hydrol.}]
			\step[fieldsource=\regexp{$MAPLOOP}, match={Internationales}, replace={Int.}]
			\step[fieldsource=\regexp{$MAPLOOP}, match={International}, replace={Int.}]
			\step[fieldsource=\regexp{$MAPLOOP}, match={Illuminating}, replace={Illum.}]
			\step[fieldsource=\regexp{$MAPLOOP}, match={Isotopes}, replace={Isot.}]
			\step[fieldsource=\regexp{$MAPLOOP}, match={Imaging}, replace={Imag.}]
			\step[fieldsource=\regexp{$MAPLOOP}, match={Israel}, replace={Isr.}]
			\step[fieldsource=\regexp{$MAPLOOP}, match={Industrial}, replace={Ind.}]
			\step[fieldsource=\regexp{$MAPLOOP}, match={Japan}, replace={Jpn.}]
			\step[fieldsource=\regexp{$MAPLOOP}, match={Information}, replace={Inf.}]
			\step[fieldsource=\regexp{$MAPLOOP}, match={Journal}, replace={J.}]
			\step[fieldsource=\regexp{$MAPLOOP}, match={Informatics}, replace={Inform.}]
			\step[fieldsource=\regexp{$MAPLOOP}, match={Knowledge}, replace={Knowl.}]
			\step[fieldsource=\regexp{$MAPLOOP}, match={Laboratory}, replace={Lab.}]
			\step[fieldsource=\regexp{$MAPLOOP}, match={Laboratories}, replace={Lab.}]
			\step[fieldsource=\regexp{$MAPLOOP}, match={Mathematical}, replace={Math.}]
			\step[fieldsource=\regexp{$MAPLOOP}, match={Language}, replace={Lang.}]
			\step[fieldsource=\regexp{$MAPLOOP}, match={Mathematics}, replace={Math.}]
			\step[fieldsource=\regexp{$MAPLOOP}, match={Abstracts}, replace={Abstr.}]
			\step[fieldsource=\regexp{$MAPLOOP}, match={Analysis}, replace={Anal.}]
			\step[fieldsource=\regexp{$MAPLOOP}, match={Academy}, replace={Acad.}]
			\step[fieldsource=\regexp{$MAPLOOP}, match={Annals}, replace={Ann.}]
			\step[fieldsource=\regexp{$MAPLOOP}, match={Accelerator}, replace={Accel.}]
			\step[fieldsource=\regexp{$MAPLOOP}, match={Annual}, replace={Annu.}]
			\step[fieldsource=\regexp{$MAPLOOP}, match={Acoustics}, replace={Acoust.}]
			\step[fieldsource=\regexp{$MAPLOOP}, match={Apparatus}, replace={App.}]
			\step[fieldsource=\regexp{$MAPLOOP}, match={Active}, replace={Act.}]
			\step[fieldsource=\regexp{$MAPLOOP}, match={Applications}, replace={Appl.}]
			\step[fieldsource=\regexp{$MAPLOOP}, match={Administration}, replace={Admin.}]
			\step[fieldsource=\regexp{$MAPLOOP}, match={Applied}, replace={Appl.}]
			\step[fieldsource=\regexp{$MAPLOOP}, match={Administrative}, replace={Administ.}]
			\step[fieldsource=\regexp{$MAPLOOP}, match={Approximate}, replace={Approx.}]
			\step[fieldsource=\regexp{$MAPLOOP}, match={Advanced}, replace={Adv.}]
			\step[fieldsource=\regexp{$MAPLOOP}, match={Advances}, replace={Adv.}]
			\step[fieldsource=\regexp{$MAPLOOP}, match={Archives}, replace={Arch.}]
			\step[fieldsource=\regexp{$MAPLOOP}, match={Archive}, replace={Arch.}]
			\step[fieldsource=\regexp{$MAPLOOP}, match={Aeronautics}, replace={Aeronaut.}]
			\step[fieldsource=\regexp{$MAPLOOP}, match={Artificial}, replace={Artif.}]
			\step[fieldsource=\regexp{$MAPLOOP}, match={Aerospace}, replace={Aerosp.}]
			\step[fieldsource=\regexp{$MAPLOOP}, match={Assembly}, replace={Assem.}]
			\step[fieldsource=\regexp{$MAPLOOP}, match={Affective}, replace={Affect.}]
			\step[fieldsource=\regexp{$MAPLOOP}, match={Association}, replace={Assoc.}]
			\step[fieldsource=\regexp{$MAPLOOP}, match={Africa}, replace={Afr.}]
			\step[fieldsource=\regexp{$MAPLOOP}, match={African}, replace={Afr.}]
			\step[fieldsource=\regexp{$MAPLOOP}, match={Astronomy}, replace={Astron.}]
			\step[fieldsource=\regexp{$MAPLOOP}, match={Aircraft}, replace={Aircr.}]
			\step[fieldsource=\regexp{$MAPLOOP}, match={Astronautics}, replace={Astronaut.}]
			\step[fieldsource=\regexp{$MAPLOOP}, match={Algebraic}, replace={Algebr.}]
			\step[fieldsource=\regexp{$MAPLOOP}, match={Astrophysics}, replace={Astrophys.}]
			\step[fieldsource=\regexp{$MAPLOOP}, match={American}, replace={Amer.}]
			\step[fieldsource=\regexp{$MAPLOOP}, match={Atmosphere}, replace={Atmos.}]
			\step[fieldsource=\regexp{$MAPLOOP}, match={Atomic}, replace={At.}]
			\step[fieldsource=\regexp{$MAPLOOP}, match={Atoms}, replace={At.}]
			\step[fieldsource=\regexp{$MAPLOOP}, match={Broadcasting}, replace={Broadcast.}]
			\step[fieldsource=\regexp{$MAPLOOP}, match={Australasian}, replace={Australas.}]
			\step[fieldsource=\regexp{$MAPLOOP}, match={Bulletin}, replace={Bull.}]
			\step[fieldsource=\regexp{$MAPLOOP}, match={Australia}, replace={Aust.}]
			\step[fieldsource=\regexp{$MAPLOOP}, match={Bureau}, replace={Bur.}]
			\step[fieldsource=\regexp{$MAPLOOP}, match={{Automatic }}, replace={Autom.}]
			\step[fieldsource=\regexp{$MAPLOOP}, match={Business}, replace={Bus.}]
			\step[fieldsource=\regexp{$MAPLOOP}, match={Automation}, replace={Automat.}]
			\step[fieldsource=\regexp{$MAPLOOP}, match={Canadian}, replace={Can.}]
			\step[fieldsource=\regexp{$MAPLOOP}, match={Automotive}, replace={Automot.}]
			\step[fieldsource=\regexp{$MAPLOOP}, match={Automobiles}, replace={Automob.}]
			\step[fieldsource=\regexp{$MAPLOOP}, match={Automobile}, replace={Automob.}]
			\step[fieldsource=\regexp{$MAPLOOP}, match={Ceramic}, replace={Ceram.}]
			\step[fieldsource=\regexp{$MAPLOOP}, match={Chemical}, replace={Chem.}]
			\step[fieldsource=\regexp{$MAPLOOP}, match={Behavioral}, replace={Behav.}]
			\step[fieldsource=\regexp{$MAPLOOP}, match={Behavior}, replace={Behav.}]
			\step[fieldsource=\regexp{$MAPLOOP}, match={Chinese}, replace={Chin.}]
			\step[fieldsource=\regexp{$MAPLOOP}, match={Belgian}, replace={Belg.}]
			\step[fieldsource=\regexp{$MAPLOOP}, match={Climatology}, replace={Climatol.}]
			\step[fieldsource=\regexp{$MAPLOOP}, match={Biochemical}, replace={Biochem.}]
			\step[fieldsource=\regexp{$MAPLOOP}, match={Clinical}, replace={Clin.}]
			\step[fieldsource=\regexp{$MAPLOOP}, match={Bioinformatics}, replace={Bioinf.}]
			\step[fieldsource=\regexp{$MAPLOOP}, match={Cognitive}, replace={Cogn.}]
			\step[fieldsource=\regexp{$MAPLOOP}, match={Biology, Biological}, replace={Biol.}]
			\step[fieldsource=\regexp{$MAPLOOP}, match={Colloquium}, replace={Colloq.}]
			\step[fieldsource=\regexp{$MAPLOOP}, match={Kolloquium}, replace={Kolloq.}]
			\step[fieldsource=\regexp{$MAPLOOP}, match={Biomedical}, replace={Biomed.}]
			\step[fieldsource=\regexp{$MAPLOOP}, match={Communications}, replace={Commun.}]
			\step[fieldsource=\regexp{$MAPLOOP}, match={Communication}, replace={Commun.}]
			\step[fieldsource=\regexp{$MAPLOOP}, match={Biophysics}, replace={Biophys.}]
			\step[fieldsource=\regexp{$MAPLOOP}, match={Compatibility}, replace={Compat.}]
			\step[fieldsource=\regexp{$MAPLOOP}, match={British}, replace={Brit.}]
			\step[fieldsource=\regexp{$MAPLOOP}, match={Components}, replace={Compon.}]
			\step[fieldsource=\regexp{$MAPLOOP}, match={Component}, replace={Compon.}]
			\step[fieldsource=\regexp{$MAPLOOP}, match={Computational}, replace={Comput.}]
			\step[fieldsource=\regexp{$MAPLOOP}, match={Delivery}, replace={Del.}]
			\step[fieldsource=\regexp{$MAPLOOP}, match={Computers}, replace={Comput.}]
			\step[fieldsource=\regexp{$MAPLOOP}, match={Computer}, replace={Comput.}]
			\step[fieldsource=\regexp{$MAPLOOP}, match={Department}, replace={Dept.}]
			\step[fieldsource=\regexp{$MAPLOOP}, match={Computing}, replace={Comput.}]
			\step[fieldsource=\regexp{$MAPLOOP}, match={Design}, replace={Des.}]
			\step[fieldsource=\regexp{$MAPLOOP}, match={Condensed}, replace={Condens.}]
			\step[fieldsource=\regexp{$MAPLOOP}, match={Detector}, replace={Detect.}]
			\step[fieldsource=\regexp{$MAPLOOP}, match={Conferences}, replace={Conf.}]
			\step[fieldsource=\regexp{$MAPLOOP}, match={Conference}, replace={Conf.}]
			\step[fieldsource=\regexp{$MAPLOOP}, match={Development}, replace={Develop.}]
			\step[fieldsource=\regexp{$MAPLOOP}, match={Congress}, replace={Congr.}]
			\step[fieldsource=\regexp{$MAPLOOP}, match={Differential}, replace={Differ.}]
			\step[fieldsource=\regexp{$MAPLOOP}, match={Consumer}, replace={Consum.}]
			\step[fieldsource=\regexp{$MAPLOOP}, match={Digest}, replace={Dig.}] 
			\step[fieldsource=\regexp{$MAPLOOP}, match={{ of the }}, replace={{ }}]
			\step[fieldsource=\regexp{$MAPLOOP}, match={{ of }}, replace={{ }}]
			\step[fieldsource=\regexp{$MAPLOOP}, match={{Of }}, replace={{ }}]
			\step[fieldsource=\regexp{$MAPLOOP}, match={{ on }}, replace={{ }}]
			\step[fieldsource=\regexp{$MAPLOOP}, match={{ On }}, replace={{ }}]
			\step[fieldsource=\regexp{$MAPLOOP}, match={{On }}, replace={{ }}]
			\step[fieldsource=\regexp{$MAPLOOP}, match={{ in }}, replace={{ }}]
			\step[fieldsource=\regexp{$MAPLOOP}, match=\regexp{\\\x{26}}, replace={{and}}] 
			\step[fieldsource=\regexp{$MAPLOOP}, match={{, and }}, replace={{, }}]
			\step[fieldsource=\regexp{$MAPLOOP}, match={{, And }}, replace={{, }}]
			\step[fieldsource=\regexp{$MAPLOOP}, match={{ and }}, replace={{ }}]
			\step[fieldsource=\regexp{$MAPLOOP}, match={{ And }}, replace={{ }}]
			\step[fieldsource=\regexp{$MAPLOOP}, match={{ In }}, replace={{ }}]
			\step[fieldsource=\regexp{$MAPLOOP}, match={{In }}, replace={{ }}]
			\step[fieldsource=\regexp{$MAPLOOP}, match={{ the }}, replace={{ }}] 
			\step[fieldsource=\regexp{$MAPLOOP}, match={{ The }}, replace={{ }}] 
			\step[fieldsource=\regexp{$MAPLOOP}, match={{The }}, replace={}] 
			\step[fieldsource=\regexp{$MAPLOOP}, match={{ for }}, replace={{ }}] 
			\step[fieldsource=\regexp{$MAPLOOP}, match={First}, replace={1st}]
			\step[fieldsource=\regexp{$MAPLOOP}, match={Second}, replace={2nd}]
			\step[fieldsource=\regexp{$MAPLOOP}, match={Third}, replace={3rd}]
			\step[fieldsource=\regexp{$MAPLOOP}, match={Fourth}, replace={4th}]
			\step[fieldsource=\regexp{$MAPLOOP}, match={Fifth}, replace={5th}]
			\step[fieldsource=\regexp{$MAPLOOP}, match={Sixth}, replace={6th}]
			\step[fieldsource=\regexp{$MAPLOOP}, match={Seventh}, replace={7th}]
			\step[fieldsource=\regexp{$MAPLOOP}, match={Eighth}, replace={8th}]
			\step[fieldsource=\regexp{$MAPLOOP}, match={Ninth}, replace={9th}]
			\step[fieldsource=\regexp{$MAPLOOP}, match={Tenth}, replace={10th}]
			\step[fieldsource=\regexp{$MAPLOOP}, match={Eleventh}, replace={11th}]
			\step[fieldsource=\regexp{$MAPLOOP}, match={Twelfth}, replace={12th}]
			\step[fieldsource=\regexp{$MAPLOOP}, match={Thirteenth}, replace={13th}]
			\step[fieldsource=\regexp{$MAPLOOP}, match={Fourteenth}, replace={14th}]
			\step[fieldsource=\regexp{$MAPLOOP}, match={Fifteenth}, replace={15th}]
			\step[fieldsource=\regexp{$MAPLOOP}, match={Sixteenth}, replace={16th}]
			\step[fieldsource=\regexp{$MAPLOOP}, match={Seventeenth}, replace={17th}]
			\step[fieldsource=\regexp{$MAPLOOP}, match={Eighteenth}, replace={18th}]
			\step[fieldsource=\regexp{$MAPLOOP}, match={Nineteenth}, replace={19th}]
			\step[fieldsource=\regexp{$MAPLOOP}, match={Twentieth}, replace={20th}]
		}
	}
}

\makeatletter
\newcommand{\linebreakand}{%
  \end{@IEEEauthorhalign}
  \hfill\mbox{}\par
  \mbox{}\hfill\begin{@IEEEauthorhalign}
}
\makeatother

\usepackage{hologo}
\usepackage{listings}
\title{Toward a Harmonized Approach -- Requirement-based Structuring of a Safety Assurance Argumentation for Automated Vehicles\\
\thanks{This work was supported by the German Federal Ministry for Economic Affairs and Climate Action within the project “Automatisierter Transport zwischen Logistikzentren auf Schnellstraßen im Level 4 (ATLAS-L4)''.}
}

\author{\IEEEauthorblockN{%
    Marvin Loba\IEEEauthorrefmark{1},
    Nayel Fabian Salem\IEEEauthorrefmark{1},
    Marcus Nolte\IEEEauthorrefmark{1},
    Andreas Dotzler\IEEEauthorrefmark{2},
    Dieter Ludwig\IEEEauthorrefmark{4},
    and Markus Maurer\IEEEauthorrefmark{1}
}\\ 
\linebreakand
\IEEEauthorblockA{%
    \IEEEauthorrefmark{1}\textit{TU Braunschweig} \\%
    \textit{Institute of Control Engineering}, Braunschweig, Germany\\%
    \{m.loba, n.salem, marcus.nolte, markus.maurer\}@tu-braunschweig.de%
}
\linebreakand
\linebreakand
\linebreakand
\IEEEauthorblockA{%
    \IEEEauthorrefmark{2}\textit{MAN Truck \& Bus SE}, Munich, Germany\\%
    andreas.dotzler@man.eu\\%
}%
\and%
\IEEEauthorblockA{%
    \IEEEauthorrefmark{4}%
    \textit{TÜV SÜD Auto Service GmbH}, Garching, Germany\\%
    Dieter.Ludwig@tuvsud.com%
}%

}%

\begin{document}

\twocolumn[
\begin{@twocolumnfalse}
\Huge {IEEE copyright notice} \\ \\
\large {\copyright\ 2025 IEEE. Personal use of this material is permitted. Permission from IEEE must be obtained for all other uses, in any current or future media, including reprinting/republishing this material for advertising or promotional purposes, creating new collective works, for resale or redistribution to servers or lists, or reuse of any copyrighted component of this work in other works.} \\ \\
{\Large Accepted to be published in \emph{2025 IEEE 28th International Conference on Intelligent
Transportation Systems (ITSC)}, Broadbeach, Australia, November 18-21, 2025.}  \\ \\ 
Cite as:\\ \\
\noindent\fbox{%
    \parbox{\textwidth}{%
        M.~Loba, N.~F.~~Salem, A.~Dotzler, D. Ludwig, and M.~Maurer, ``Toward a Harmonized Approach -- Requirement-based Structuring of a Safety Assurance Argumentation for Automated Vehicles,'' in \emph{2025 IEEE 28th International Conference on Intelligent
Transportation Systems (ITSC)}, Broadbeach, Australia, November 18-21, 2025, {to be published}.
    }%
}
\vspace{2cm}

\end{@twocolumnfalse}
]

\noindent\begin{minipage}{\textwidth}

\hologo{BibTeX}:
\footnotesize
\begin{lstlisting}[frame=single]
@inproceedings{loba_2025,
        author={{Loba}, Marvin and {Salem}, Nayel Fabian and {Nolte}, Marcus and {Dotzler}, Andreas and {Ludwig}, Dieter and {Maurer}, Markus},
        booktitle={2025 28th {International} {Conference} on {Intelligent} {Transportation} {Systems} ({ITSC})},
        title={{Toward} a {Harmonized} {Approach} -- {Requirement}-based {Structuring} of a {Safety} {Assurance} {Argumentation} for {Automated} {Vehicles}},
        address = {Broadbeach, Australia},
        year={2025},
        publisher={IEEE, to be published}
}
\end{lstlisting}
\end{minipage}

\maketitle
\thispagestyle{empty}
\pagestyle{empty}



\begin{abstract}

Despite the increasing testing operations of automated vehicles on public roads, media reports on incidents show that safety issues caused by automated driving systems persist to this day.
Manufacturers face high development uncertainty when aiming to deploy these systems in an open context. 
In particular, one challenge is establishing a valid argument at design time that the vehicles will exhibit reasonable residual risk when operating in its intended operational design domain.
While there is extensive literature on assurance cases for safety-critical systems in general, the domain of automated driving lacks explicit requirements regarding the creation of safety assurance argumentations for automated vehicles. 
In this paper, we aim to narrow this gap by elaborating a requirement-based approach. 
We identify structural requirements for an argumentation based on published literature and supplement these with structural requirements derived from stakeholder concerns. 
We apply these requirements to obtain a proposal for a generic \mbox{argumentation} structure.
The resulting ``safety arguments'' address the developed product (product argument), the underlying process (process argument) including its conformance/compliance to standards/laws (conformance/compliance argument), as well as an argumentation's context (context argument) and soundness (soundness argument). 
Finally, we outline argumentation principles in accordance with domain-specific needs and concepts.

\end{abstract}

\begin{IEEEkeywords}
safety argumentation, automated vehicles
\end{IEEEkeywords}


\section{INTRODUCTION}
\label{sec:intro}

In recent years, the testing operations of automated vehicles have advanced steadily on public roads, with growing fleets and expanding operational design domains. 
Consequently, the question arises as to why automated road vehicles have not yet been commercialized on a large scale.

One reason lies in the uncertainty surrounding the deployment of such systems in an open context.\footnote{Refers to an environment that cannot be fully specified at design time, either due to its complexity, unpredictability, or temporal development~\cite{BurtonHawkins_2020}.}
During operation, automated vehicles are exposed to various kinds of uncertainty, e.g., regarding measurements or the prediction of the behavior of other road users.
Knowledge gaps are inevitable, resulting in an incomplete specification of requirements, which, in turn, culminates in incomplete testing.
Such functional and systemic causes lead to an inherent risk to all participants in the traffic system.
This inherent risk, which is posed by the operation of automated vehicles, can be reduced by conscious development but never eliminated~\cite{maurer2018,Nolte2021}.

As the complexity of safety assurance scales with these effects of uncertainty, the established practice of simply consolidating evidence stemming from activities in the safety lifecycle is no longer sufficient to release automated vehicles.
Instead, there is a need for a coherent argument that addresses how the absence of \emph{unreasonable risk}\footnote{``Risk judged to be unacceptable in a certain context according to valid societal moral concepts''~\cite[Part~1,~3.176]{ISO26262_2018}.} is achieved and how valid the associated assumptions remain during field operation.

Frequently also referred to as a ``safety case'' (see~\cref{subec:terminology} for a terminological delimitation), one common approach to respond to this task is a ``safety assurance argumentation.''
Crafting such an artifact is expected by regulation \cite[]{eu1426}~and standards \cite{UL4600_2023,ISO8800_2024,ISOIECIEEE15026_2_2022,ISO26262_2018}.
Although it is possible to realize argumentations at different levels of formalization, a semi-formal representation (e.g., by using the Goal Structuring Notation (GSN), see~\cite{GSN_Standard_2021}; originally proposed by Kelly~\cite{Kelly_1998}) appears to be a suitable compromise, as a textual degree of freedom is sustained while concepts like hierarchy and modularity support the management of complexity.

Safety assurance argumentations for complex systems have been comprehensively researched and addressed in the literature for decades~\cite{Kelly_1998,Kelly2018,Hawkins2011,Hawkins_2025,Hawkins2022}.
Nonetheless, although extensive literature deals with safety assurance argumentations, the published state of the art lacks an explicit provision of structural requirements for the design and verification of a GSN-based safety assurance argumentation for automated vehicles.

In this paper, we aim to narrow this gap by providing such requirements.
For this, we first analyze terminological inconsistencies in the context of safety assurance argumentations.
We address them by providing an ontology (\Cref{subec:terminology}), as the harmonization of terminology and concepts facilitates stakeholder communication and supports the applicability of requirements specified in this paper. 
Second, in \Cref{sec:reqID}, we derive structural requirements for the creation of a safety assurance argumentation based on a comprehensive literature review (see~\Cref{subsec:RW}) and supplement these by requirements derived from identified stakeholder concerns.
Third, we propose a generic argumentation structure that satisfies the specified requirements (\Cref{sec:argumentationapproach}).
Additionally, we aim to demonstrate the basic applicability of this requirement-based argumentation structure to the domain of automated driving.
Hence, we outline central argumentation contents from a GSN-based safety assurance argumentation that follows the presented framework and underlies this paper.
Finally, we discuss the open issues related to the presented approach.


\section{BACKGROUND}
\label{sec:background}

\subsection{Terminology}
\label{subec:terminology}

\begin{figure*}
	\centering
    \includegraphics[width=0.725\textwidth]{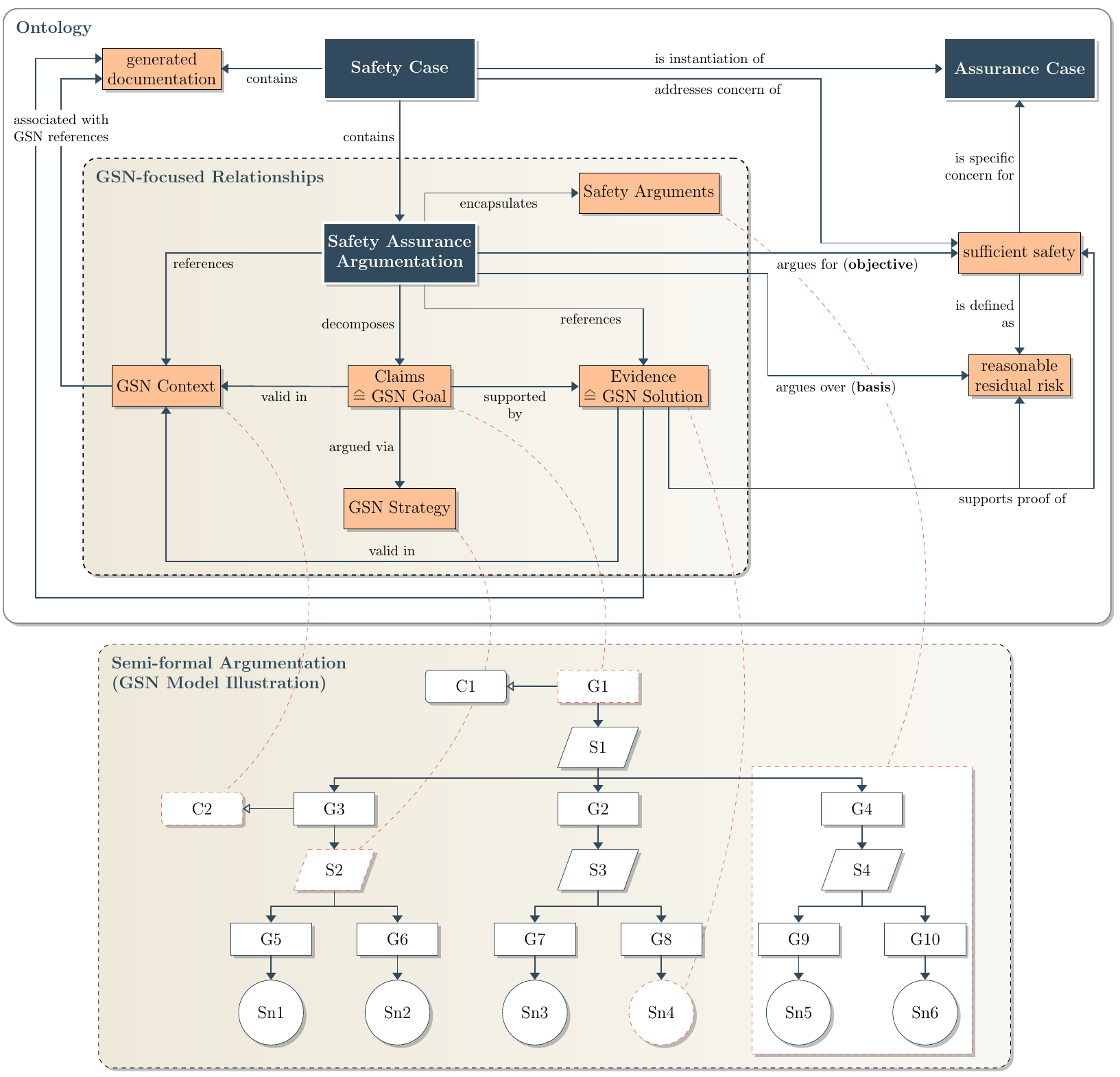}
	\caption[]{Proposed ontology in the context of safety cases.~\ArtifactLegend~and \OntologyLegend~indicate artifacts and other ontology elements. 
    GSN goals, strategies, contextual artifacts, and solutions are represented by \GoalLegend,~\StrategyLegend,~\ContextLegend,~and \SolutionLegend,~respectively.
    }
	\label{fig:terminology} 
\end{figure*}

The terms ``safety case'' and ``safety assurance argumentation'' are often used interchangeably.
A conceptual distinction is illustrated in \Cref{fig:terminology} to clarify on their relationship.

The overarching concept is an ``assurance case,'' defined as an ``auditable artifact that provides a convincing and sound argument for a claim on the basis of tangible evidence under a given context''~\cite[\nopp 3.1.1]{ISOIECIEEE15026_2_2022}.
While the principles of an assurance case apply equally for different properties of a complex system whose proof is pursued~\cite{ACWG_Guidance_2021}, the specific concern for a safety case is the emergent property \emph{safety}.
Hence, the latter can be understood as a dedicated instantiation of an assurance case.

Multiple standardized definitions (e.g., \cite[\nopp 4.2.37]{UL4600_2023}, \cite[\nopp 3.15]{BSI1881_2022}, and \cite[\nopp 3.136]{ISO26262_2018})
exist that share certain characteristics attributed to a safety case.
Correspondingly, a structured \textit{argument} that is supported by \textit{evidence} and considered in a specific environment (\textit{context}) shall prove \textit{safety}.
Accordingly, \Cref{fig:terminology} visualizes that evidence supports the claim of sufficient safety as the argumentation objective. 
Nevertheless, as safety is defined as absence of unreasonable risk in the context of road vehicles~\cite[Part 1,~3.132]{ISO26262_2018}, the basis of the argumentation relies on residual risk.
Implications for the starting point of the argumentation are discussed in \Cref{subsec:TL_Claim}.

Distinguishing a ``safety assurance argumentation'' from a ``safety case'' emphasizes the particular task of building a coherent argumentation that goes beyond consolidating evidence generated during safety assurance processes.\footnote{This perspective is supported by requirements defined in the recently published ISO~PAS~8800:2024~\cite[\nopp 7.3.4 e)]{ISO8800_2024}.}
Instead, a dedicated argumentation artifact is required that demonstrates the contributions of documented work products to achieve the absence of unreasonable risk.
This is pursued by systematically decomposing claims using strategies and references to evidence and context.
The modeled claims, as well the evidences, are valid in a specific context --- see \cite{Palin_2010} for a discussion of context dependency in automotive safety arguments.

Thus, the safety case comprises the safety assurance argumentation, which in this paper is understood as a GSN-based model, as well as the documentation associated with evidence and context elements referenced within the argumentation.\footnote{This interpretation is shared by~\cite[\nopp 1]{GraydonLehman_LLM_Arguments2025}. 
Accordingly, a ``safety argument'' forms a ``safety case'' once it is considered together with the ``materials it references.''
However, we recommend the usage of ``safety argumentation'' instead of ``safety argument,'' as the latter is a common label for a distinct branch within an argumentation (see \cites{Hawkins_2025}{Hawkins2011}).}
 The lower part of  \Cref{fig:terminology} visualizes correspondences between GSN elements in the ontology and their exemplary use in a schematic GSN model.


\subsection{Related Work}
\label{subsec:RW}



Structure and content of assurance cases are covered by standards~\cite{ISOIECIEEE15026_2_2022}, best practices~\cite{ACWG_Guidance_2021}, and publications that provide the required ``tools'' like GSN~\cite{GSN_Standard_2021}.
Comprehensive guidance for the development of safety cases is available that provides methodological approaches for responding to common pitfalls and challenges~\cite{Kelly_1998,Hawkins_2025,Hawkins2022}.

\cite{Birch2014,Hawkins2011,Hawkins_2025,Habli2006,Fenn2024} address the task of structuring a safety case, especially with the support of differentiating safety argument types, such as risk, confidence, and operational arguments.
\textcite{Birch2014} propose a layered approach for safety argumentations as an adaption of the risk/confidence argument approach, emphasizing the necessity for conceptualizing a structured approach to create safety argumentations. 

The standard UL~4600 supports the sufficiency of a claim-based safety case for ``autonomous systems,'' as the standard ``puts forth assessment criteria to determine the acceptability of a safety case''~\cite[\nopp 1.2.3]{UL4600_2023}. 
However, it neither presents a process nor requirements for constructing an argumentation.

Domain-specific literature on safety argumentations for road vehicles includes, e.g., work in the context of functional safety~\cite{Palin_2010,Horiba2019,ISO26262_2018} or safety of artificial intelligence~\cite{ISO8800_2024}.
Furthermore, manufacturers often disclose to the public their safety assurance approaches for automated vehicles, for example, through text-based safety reports.
Unfortunately, a coherent line of argumentation is frequently not apparent with these representations and is merely implied.
However, publications specifically detailing safety case approaches (e.g.,~\cite{Favaro2023}) may be considered to inform the structuring of a GSN-based argumentation.
Another example would be Aurora~\cite{aurora2023safetycase} providing an argumentation representation oriented toward GSN, revealing a superordinate argumentation structure; however, it is tailored for external stakeholder communication, as, for instance, no evidence is provided to the claims made.

Different frameworks aim to support the creation of GSN-based safety argumentations for automated vehicles~\cite{reich2023,Wagner2024,Kodama2024}.
Still, these lack traceability to requirements, i.e., they miss an evident reasoning for the resulting argumentation structure.
Since the references do not explicitly address the formulation of structural requirements, we aim to specify requirements that may be implicitly captured in the cited work and, if necessary, supplement them with additional requirements.


\section{REQUIREMENTS}
\label{sec:reqID}

In the following, relevant literature is examined to derive macro- and microstructural requirements in \Cref{subsec:LitReqs}.
The former refers to requirements toward the superordinate structure, i.e., the distinction of individual safety arguments.
The latter refers to requirements toward the subordinate structure, i.e., the specific contents that shall be covered in the downstream argumentation contained in the safety arguments.
Based on our joint experience in creating and assessing safety argumentations, stakeholder concerns are identified and translated into supplementary structural requirements in \Cref{subsec:AddReq}.

\subsection{Literature-based Requirement Derivation}
\label{subsec:LitReqs}

Following Hawkins et al.~\cite[p.~6]{Hawkins2011}, we argue that a clear distinction between ``risk arguments'' and ``confidence arguments'' is a key factor to providing compelling safety argumentations.
Risk arguments shall capture the direct causal chain of risk mitigation (\MacroReqText{1}), whereas confidence arguments shall support the confidence in the risk argument~\cite{Hawkins_2025,Kelly2018}.

Assurance Claim Points have been introduced in~\cite{Hawkins2011,GSN_Standard_2021} to explicitly capture this relationship and indicate the assertions in a risk argument whose adequacy is argued over in separate confidence arguments.
Hence, there are fragments to the ``overall confidence argument'' (\MacroReqText{2}) distributed within the safety argumentation~\cite{Hawkins2011}.
Assurance Claim Points are also used in automotive safety arguments~\cite{Horiba2019}.

Kelly~\cite{Kelly2018} introduces the ``conformance/compliance argument'' as an additional safety argument type that argues over adherence to relevant standards, regulations, and legislation.
In~\cite[2:5.2.1]{ACWG_Guidance_2021} the categorization via aforementioned argumentation types is adopted, but the authors use the label ``conformance argument'' only.
While the literature often refers to compliance with standards (e.g.,~\cite{Wagner2024,Kodama2024}), conformance is defined as ``voluntary adherence to a standard, specification,  guide, process or practice'' and compliance as ``forced adherence to a law, regulation, rule or process''~\cite{ACWG_Guidance_2021}.
This is also in accordance with the distinction made in~\cite{Swaminathan2025}.
Hence, contrary to the assumption in \cite{ACWG_Guidance_2021} that compliance subsumes under conformance, we deem distinguishing the two dimensions helpful.
This is due to regulatory requirements being mandatory, whereas arguing for conformance includes an upstream identification of relevant normative requirements.
It is worth noting that both concepts are closely linked and potentially challenging to separate, especially since regulation may demand adherence to standards, which converts the corresponding normative into mandatory requirements. 

However, practical experience has shown that the distinction promotes the achievement of a separation of concerns, fostering clarity in stakeholder communication.
Thus, we propose a conformance argument (\MacroReqText{3}) and a compliance argument (\MacroReqText{4}) that encapsulate arguments that the development adheres to normative and regulatory requirements, respectively. 

Arguing ``safety through direct appeal to features of the implemented item'' is often termed a product argument.
Arguing through ``appeal to features of the development and assessment process'' is often termed a process argument~\cite[Part~10,~5.3.1]{ISO26262_2018}.
This classification is supported by other ISO documents~\cite[\nopp 8.5.1]{ISO8800_2024} as well as research~\cite{Luo2016,reich2023,Habli2006}, leading to
\MacroReqText{5}.

The preceding explanations yield following requirements:

\begin{tcolorbox}[colback=RedHigh!2, colframe=rotx, sharp corners=southwest, rounded corners=southeast, boxrule=0.8mm, left=1mm, right=1mm, top=1mm, bottom=1mm,title=Macrostructural Requirements,fonttitle=\fon{pbk}\scshape]
	The superordinate safety argumentation shall include a:
	\begin{itemize}
		\setlength{\itemsep}{.1em}
		\item[\MacrostructReq] \textbf{\textcolor{rotx}{risk argument}} that argues over risk reduction. \MacroReq{1}
		\item[\MacrostructReq] distributed overall \textbf{\textcolor{rotx}{confidence argument}} that argues why elements or their assertion in the risk argument should be trusted. \MacroReq{2}
		\item[\MacrostructReq] \textbf{\textcolor{rotx}{compliance argument}} that argues for adherence to regulatory requirements. \MacroReq{3}
		\item[\MacrostructReq] \textbf{\textcolor{rotx}{conformance argument}} that argues for adherence to normative requirements. \MacroReq{4}
		\item[\MacrostructReq] risk argument comprising a \textbf{\textcolor{rotx}{product argument}} and a \textbf{\textcolor{rotx}{process argument}}. \MacroReq{5}
	\end{itemize}
\end{tcolorbox}

While the overarching goal of the risk argument is to argue over risk management, Kelly emphasizes that this is directly related to arguing over the appropriate management of hazards~\cite{Kelly_1998}.
The corresponding argument encompasses the elimination or mitigation of all identified hazards posed by the system as well as linking it to the resulting risk.
Similarly, Hawkins et al. highlight that ``everything that is included as part of a risk argument must have a direct role as part of the causal chain to the hazard''~\cite{Hawkins_2025}, consequently yielding \MicroReqText{6}.

Palin and Habli~\cite[Fig.~3]{Palin_2010} consider a ``Through Life Safety Argument'' as part of the ``High Level Vehicle Safety Argument Pattern'' they present --- marking another requirement that emerges from the demand to account for the operational phase, i.e., to argue over the whole system lifecycle (\MicroReqText{7}).
This concern also becomes evident in~\cite{Hawkins2022}, as the authors extend the top-level claim of sufficient safety by the notion of ``throughout its entire operational life.''\footnote{In this regard, Fenn et al.~\cite{Fenn2024} extend the concept of Assurance Claim Points by introducing ``Operational Claim Points'' to allow for establishing operational arguments that can be understood as a runtime-focused perspective associated with the risk argument.}

Wagner and Carlan incorporate the claim that the developing organization is trustworthy in the superordinate structure of their argumentation framework~\cite{Wagner2024}, positioning it alongside the risk argument.
This consideration is related to arguing over the implementation of a safety culture and is also addressed by UL~4600~\cite{UL4600_2023} as well as Aurora~\cite{aurora2023safetycase}.
This aspect is captured via \MicroReqText{8}.

The preceding explanations yield following requirements:

\begin{tcolorbox}[colback=BlauDark!2, colframe=BlauDark, sharp corners=southwest, rounded corners=southeast, boxrule=0.8mm, left=1mm, right=1mm, top=1mm, bottom=1mm,title=Microstructural Requirements,fonttitle=\fon{pbk}\scshape]
	The subordinate safety argumentation shall argue over:
	\begin{itemize}
		\setlength{\itemsep}{.1em}
		\item[\MicrostructReq] hazards posed by a system and discuss how these \textbf{\textcolor{BlauDark}{hazards are managed}} by adequate measures. \MicroReq{6}
		\item[\MicrostructReq]  \textbf{\textcolor{BlauDark}{system lifecycle}} considerations, including operational aspects related to post-deployment activities. \MicroReq{7}
		\item[\MicrostructReq] how the process accounts for both procedural and underlying organizational aspects, such as the establishment of a \textbf{\textcolor{BlauDark}{safety culture}}.  \MicroReq{8}
	\end{itemize}
\end{tcolorbox}

\subsection{Additional Requirements Based on Stakeholder Concerns}
\label{subsec:AddReq}


While we elicited the macro- and microstructural requirements based on the identification of common principles we found in the literature, the need for additional requirements arises when stakeholder concerns are considered.
Internal stakeholders (e.g., function developers, managers, or safety engineers) involved in the creation of the argumentation often possess implicit knowledge that enables them to comprehend all aspects of the argumentation.
To enable conscious assessments by external stakeholders, such as audits by certification agencies or type approval authorities, we encourage making this knowledge explicit. 
This intention is especially tied to the objective of achieving a safety argumentation structure that is as self-explanatory as possible.

\textcolor{DeepGreen}{\AddReqText{9}~\textbf{Contextualization Argument}}\quad 
The objective of making associated knowledge explicit necessitates a sufficient contextualization of the argumentation objective, providing sufficient context that, in turn, establishes an adequate argumentation basis for the downstream argumentation complexes.
This contextualization can be understood as an ``onboarding'' of external stakeholders. 
From our experience, implicit knowledge can be associated with individual concepts, terminology, and abbreviations leveraged by an organization when creating the argumentation.
Complementary to this, we deem a basic contextualization of the system of interest and its operation as important context dimensions, ideally encapsulated in a dedicated contextualization argument.

\textcolor{DeepGreen}{\AddReqText{10}~\textbf{Soundness Argument}}\quad Additionally, we propose a soundness argument that argues over different measures to account for uncertainty.
We consider an argument to be ``sound'' if domain experts judge that the remaining uncertainty from an argumentation has been sufficiently mitigated. 
In the context of a safety assurance argumentation, various sources of uncertainty exist, including uncertainty regarding the validity of the claims' inference, the scope and relevance of context, as well as the relevance and the validity of evidence~\cite[\nopp 4.1]{ISOIECIEEE15026_2_2022}.

To enable comprehension by external stakeholders, the soundness argument shall argue over all applied methods that were used in the course of creating and maintaining the argumentation in order to ensure its soundness.
As already introduced, Assurance Claim Points can be utilized to reduce uncertainty in the appropriateness of GSN elements and their assertions within a graphical argument.
Hence, the soundness argument may include the reasoning of the ``overall confidence argument'' (see~\cite{Hawkins_2025}).
Such a reasoning should provide insights on both the selection of elements in the risk argument that are associated with Assurance Claim Points as well as explanations on how the aggregation of Assurance Claim Points purposefully contributes to an overall satisfactory level of confidence.
Other measures include for example independent reviews or methods to identify and manage weaknesses (e.g., by using \emph{challenges} and \emph{defeaters}) as complementary steps to developing a risk argument~\cite{Hawkins_2025}.
As an example for quantitative assessments, Herd and Burton propose the use of Subjective Logic to propagate uncertainty in GSN-based argumentations~\cite{Herd2024}.


The preceding explanations yield following requirements:
\begin{tcolorbox}[colback=DeepGreen!2, colframe=DeepGreen, sharp corners=southwest, rounded corners=southeast, boxrule=0.8mm, left=1mm, right=1mm, top=1mm, bottom=1mm,title=Supplementary Requirements,fonttitle=\fon{pbk}\scshape]
	The superordinate safety argumentation shall include a:
	\begin{itemize}
		\setlength{\itemsep}{.1em}
		\item[\AdditionalReq] \textcolor{DeepGreen}{\textbf{contextualization argument}} addressing relevant context dimensions to allow for comprehension of the downstream argumentation. \AddReq{9}
		\item[\AdditionalReq] \textcolor{DeepGreen}{\textbf{soundness argument}} that argues over applied methods to account for uncertainty in the argumentation's overall validity. \AddReq{10}
	\end{itemize}
\end{tcolorbox}


\section{ARGUMENTATION APPROACH}
\label{sec:argumentationapproach}

\begin{figure*}[h]
	\centering
	\includegraphics[width=0.72\textwidth]{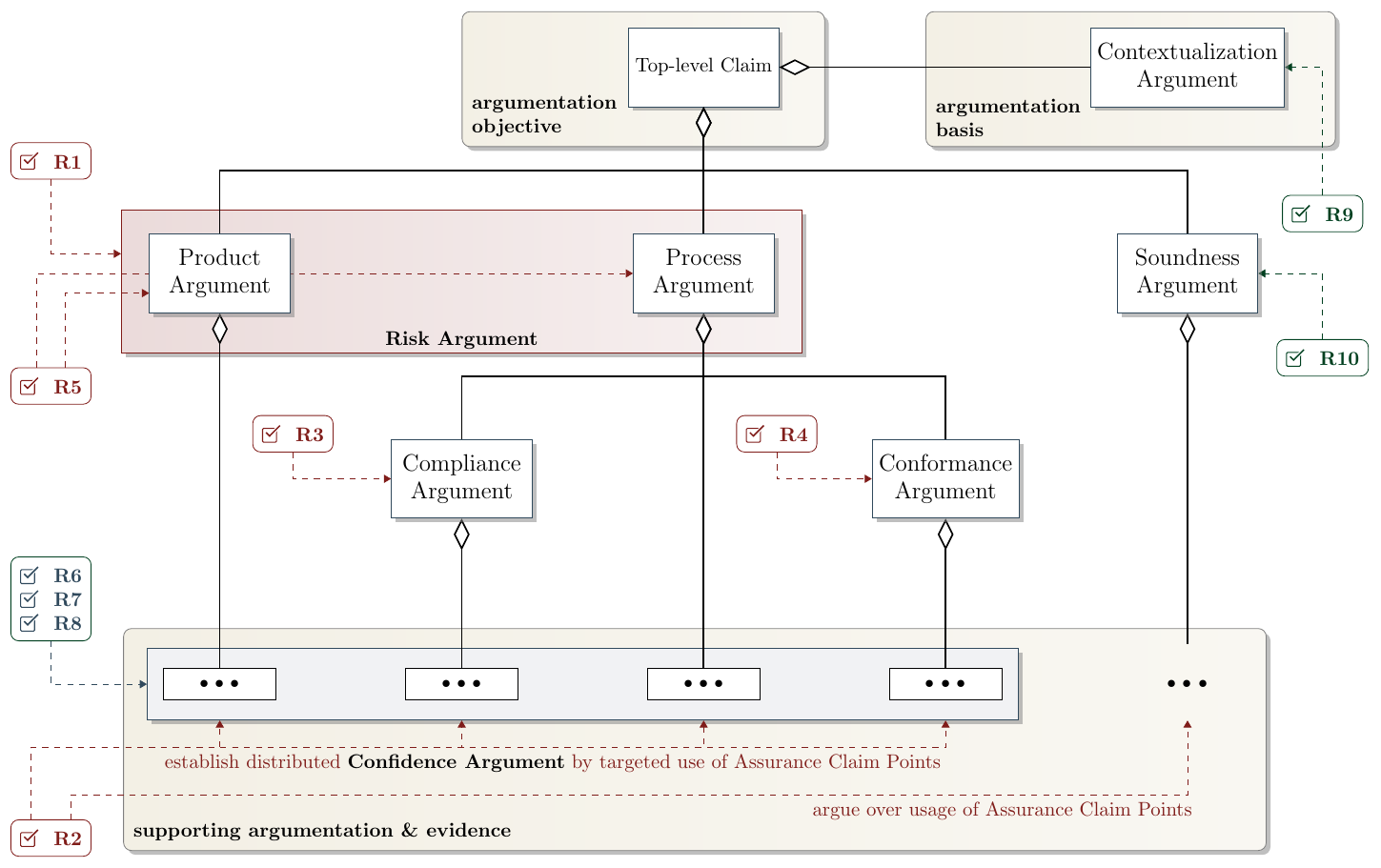}
	\caption[]{Conceptual illustration for structuring a safety assurance argumentation based on the implementation of the requirements specified in this paper. 
		The satisfy relationship (\SatisfyLegend) represents the allocation of elicited requirements, with macrostructural and supplementary requirements posing the primary basis for distinguishing the safety arguments.
		The aggregation relationship (\AggregationLegend) visualizes hierarchical decomposition.
		The microstructural requirements are associated with the supporting argumentation contained in both the product and process argument.}
	\label{fig:arg_overview}
\end{figure*}

\Cref{fig:arg_overview} illustrates our proposed structure of a safety assurance argumentation that satisfies the defined requirements.
In the remainder of this section, we will describe key argumentation principles\footnote{Measures argued over in the soundness argument are agnostic to the technology since the methods considered for dealing with uncertainties are argumentation-theoretical and apply the same for different system contexts.
Hence, we refrain from discussing the soundness argument's contents in depth.
} 
of the framework's instantiation to demonstrate its general applicability with respect to concepts established in the field of automated driving.
The line of argumentation follows an underlying GSN model developed in the \mbox{ATLAS-L4} project and primarily oriented toward the argumentation framework of the VVMethods project in~\cite{reich2023}.
However, several aspects of the VVMethods argumentation framework were adapted or extended in the project to account for all requirements specified in this paper.
This includes, e.g., introducing a contextualization and soundness argument, explicitly addressing conformity, or distinguishing risk acceptance criteria regarding their abstraction level, as discussed in this section.

\subsection{Top-level Claim}
\label{subsec:TL_Claim}

The claim of a system being safe 
has to be accompanied by a definition of what constitutes safe operation, as suggested in~\cite{Hawkins2011,Palin_2010,Wagner2024,BurtonHawkins_2020,Hawkins_2025}.
The need for justifying the top-level claim is also formulated as a normative requirement, linked to comments on this justification's critical character since it ``drives the assurance case's formulation'' and ``serves as a means for communicating''~\cite{IEEE15026_2_2011}.

Following Fleischer~\cite{Fleischer2023}, we argue that, from a linguistic point of view, safety is an ``open signifier.''
This means that the usage of the term ``safety'' both enables but also impedes interdisciplinary communication.
This is due to the term's openness.
Accordingly, while there is an alleged consensus among stakeholders on the objective of deploying ``safe'' automated vehicles, implicit and deviating stakeholder understandings, e.g., regarding society's and engineers' interpretations, exist when it comes to the definition of safety.
The range of stakeholder perspectives on safety and risk in the field is also discussed by Salem et al.~\cite{Salem2023}.

From an engineering perspective, there is far-reaching consensus in the domain of automated driving that safety is defined as absence of unreasonable risk (see~\cite{ISO21448_2022,ISO26262_2018,Favaro2023}).
This definition acknowledges that inherent risk prevents achieving freedom from risk during operation.
In line with~\cite{Nolte_unpublished}, we deem it especially important to avoid unfulfillable stakeholder expectations of ``zero risk'' (associated with a ``Vision Zero'') by explicitly representing and communicating residual risk.

Correspondingly, we consider the \emph{absence of unreasonable risk} as a favorable top-level claim.
This approach is also taken in the literature, both within the domain (e.g.,~\cites{Kodama2024,reich2023}) but also for assurance cases in general, albeit with slightly different wording (``no intolerable risk'' according to~\cite{ACWG_Guidance_2021}).
To still account for stakeholder expectations and facilitate communication by using established labels, e.g., with respect to an assessor's aim to assess whether the system is ``safe'' when reviewing a safety case, we propose the contextualization argument as a possibility to allocate further explanation how the concepts of residual risk and safety relate.

\subsection{Contextualization Argument}
\label{subsec:ContextualizationArg}


Despite the aforementioned potential to argue over the justification of the top-level claim by defining safety via risk or, conversely, relating risk to safety, content dimensions must be contextualized so that external stakeholders can comprehend the argumentation.
For instance, a system's definition and a description of its operating role and environment can pose top-level contextual elements~\cite[]{ISOIECIEEE15026_2_2022}.
We regard the following documentation as highly relevant for automated vehicles:
\begin{itemize}
	\item Operational Concept according to~\cite{ISOIECIEEE15288_2023}, including
	\begin{itemize}
		\item Operational Design Domain (cf.~\cite{ISO34503_2023})
		\item Behavior specification/competencies (cf.~\cites{Salem2024_BehSpec,BSI1891_2025})
	\end{itemize}
    \item Concept of Operations according to~\cite{ISOIECIEEE15288_2023}
	\item System Description (cf.~\cites{ISO26262_2018,ISO21448_2022})
	\item Concept explanations (e.g., the introduced \emph{inherent risk})
\end{itemize}

\subsection{Process Argument} 
\label{subsec:ProcessArg}

The process argument addresses aspects that contribute to determining whether the organization is capable of developing an automated vehicle that is free from unreasonable risk.
This comprises covering cultural aspects.
In this regard, the argumentation addresses the establishment of a safety culture~\cite[Part~1,~3.137]{ISO26262_2018}
within an organization, e.g., by arguing over safety policies or safety-related trainings and procedures for onboarding employees~(\MicroReqText{8}).

Furthermore, this refers to arguing over the development process, including post-deployment activities (cf. \cite{ISOIECIEEE15288_2023}).
The argumentation needs to provide an adequate information basis regarding the definition and assessment of relevant sub-processes in subsequent phases as well as proof for the deployment of these processes.
This proof may be provided by evidence emerging from conducted reviews which attest that defined processes are being practiced.

The distinction between sub-processes can be derived from technical processes that subsume under the lifecycle processes consistent with system engineering standards~\cite{ISOIECIEEE15288_2023}.
The lifecycle perspective is not only accounted for by the associated operation and maintenance sub-processes that define post-deployment activities (\MicroReqText{7}) but also by arguing that the processes are scrutinized and, in case of identified deficiencies, adapted in order to achieve continuous improvement.

One principal aspect factoring into the assessment of the processes' suitability is the adherence to normative and regulatory requirements.
Therefore, the process definition is supported by the adjacent conformity and compliance arguments.\footnote{Subordinating the conformity and compliance argument to the process argument and linking it via the claim of the appropriateness of the processes following the state of the art is different from the referenced literature, which separates these from the process argument.}

\subsubsection{Conformity Argument}
\label{subsubsec:ConformityArg}

Even if the codification of the state of the art is one of the objectives of standardization, there is no agreed-upon state of the art that prescribes which normative documents are to be taken into account when developing automated vehicles.
This situation is made particularly difficult by the fact that the normative landscape is dynamic.
Normative documents, which exhibit varying degrees of maturity and present both complementary and competing approaches, currently are published at high frequency~\cite{Nolte_unpublished}. 
Therefore, a critical (see also~\cite[\nopp 2.1.3]{KoopmanArgumentation_2019}) analysis is required that provides a rationale for selecting normative documents, i.e., for gathering the relevant normative requirements that determine the definition of the development process. 

As the analysis of standards involves multiple assumptions, it is crucial to guarantee traceability within the argumentation.
This traceability shall be established between normative requirements associated with the underlying standards in the conformity argument and the resulting decisions for the process design argued over in the process argument.
As Kelly~\cite{Kelly2018} explains, there should be an overlap between the conformance argument and the risk argument. 

\subsubsection{Compliance Argument}
\label{subsubsec:ComplianceArg}

The compliance argument follows argumentation principles that are comparable to those of the conformity argument.
However, arguing for adherence to regulatory requirements demands an ex-ante translation into engineering requirements in the first place.
This task is especially difficult, as legal texts are often open to interpretation.
There is still a lack of court rulings that provide practical interpretations of legal clauses in the context of automated vehicles.
Additionally, as discussed in~\cites{Nolte_unpublished2,Nolte2025AFGBV}, challenges are present due to differences in the conceptualization of \emph{safety} in different legal frameworks.

\subsection{Product Argument}
\label{subsec:ProductArg}

While the process argument provides evidence for the organization's capabilities, the product argument provides evidence that the vehicle possesses the capability not to pose an unreasonable risk when operating in its operational design domain.
The main argumentation principle supporting this claim is the fulfillment of stakeholder-dependent risk acceptance criteria, i.e., the system satisfies specified risk thresholds.

To this end, we propose distinguishing between ``global'' and ``scenario-based'' risk acceptance criteria.
A similar delimitation of complementary perspectives is presented in~\cite{Favaro2023,Kodama2024}.
The global perspective refers to a scenario-independent evaluation of the aggregated system performance by statistical means.
This requires gathering data during the automated vehicle's operation in its designated operating environment.
In contrast, scenario-based acceptance criteria correspond to a scenario-based risk evaluation.

From an argumentation standpoint, both argumentation strands follow the same pattern:
Acceptance criteria of the respective abstraction level need to be defined in accordance with stakeholder expectations, evaluated to be met, and be maintained.
Arguing for maintenance is associated with conducting field operation, gathering evidence, and ensuring that safety-related incidents do not violate the criteria after deployment.

In terms of scenario-based acceptance criteria, in line with ISO~21448~\cite{ISO21448_2022}, we argue over residual risk in known and unknown scenarios the vehicle might encounter during its operation.
On the one hand, sufficient confidence needs to be established that residual risk in unknown scenarios will not result in the violation of any acceptance criteria.
On the other hand, risk reduction in known hazardous scenarios must be carried out sufficiently.
This involves estimating the actual risk posed by the vehicle, specifying the tolerable risk target, and implementing safety measures to iteratively reduce the risk until it is at least reduced to a tolerable level for the respective scenarios under consideration.
Following~\cite{Salem_RMC_2024}, the former two activities relate to risk assessment and the latter corresponds to risk treatment.
The argumentation dealing with the risk treatment relies on a safety concept that encompasses safety requirements and derived measures, thereby yielding the argument that all identified hazards are sufficiently mitigated or eliminated, as suggested by the literature (\MicroReqText{6}).


\section{CONCLUSION AND FUTURE WORK}
\label{sec:Conclusion}

In this paper, we contributed to overcoming challenges related to creating a safety assurance argumentation for automated vehicles.
To this end, we first proposed an ontology that distinguishes between the artifacts ``safety case'' and ``safety assurance argumentation,'' connecting them with relevant concepts and GSN model elements.
Thereby, we aim to facilitate stakeholder communication by providing a harmonized terminology that dismantles inconsistencies.

Second, we derived requirements for structuring a safety assurance argumentation based on commonalities and differences in relevant literature.
We defined supplementary requirements as a result of considering stakeholder concerns derived from our experience.
We implemented all requirements to obtain a generic requirement-based argumentation structure.

Third, we instantiated the resulting structure based on domain-specific principles, i.e., presented the core argumentation principles of a detailed GSN model underlying this paper.

While the state of the art for safety assurance processes is not explicitly defined, normative documents capture respective requirements.
In contrast, the field lacks standardization in terms of informing the creation of GSN-based safety assurance argumentations.
We deem a harmonized requirement-based approach valuable to promote consistency in argumentations.

However, the structure of arguments is by nature always characterized by subjectivity.
To account for associated uncertainty, we particularly emphasize the relevance of making assumptions in the argumentation as well as underlying knowledge explicit.
Thus, the introduced ``soundness argument'' and ``contextualization argument'' can pose important concepts that require further research, e.g., with respect to the questions of how to adequately represent evidence uncertainty or how beneficial contextualization can be achieved.

Regarding the various stakeholders affected by the development and deployment of automated vehicles, one research area we plan to investigate in the future is the manifestation of assurance cases. 
It might be reasonable to have a ``core assurance case model'' that addresses basic argumentation principles applicable to different properties --- and derive views for different stakeholders and their concerns, such as a conformity or a compliance case for certification agencies or legal stakeholders, respectively.
The idea of having multiple assurance cases for a system whose selection is based on needs and characteristics of different audiences is also supported in~\cite[\nopp 4.1]{ISOIECIEEE15026_2_2022}

As emphasized by Nolte et al. in \cite{Nolte2025AFGBV}, addressing value conflicts such as the trade-off between mobility and physical well-being is decisive when aiming to achieve public acceptance of automated vehicles.
The discussed argumentation allows for considering different dimensions of harm, e.g., the harm to mobility.
With risk being defined as a ``combination of the probability of occurrence of harm and the severity of that harm''~\cite[3.128]{ISO26262_2018}, the concept of stakeholder-dependent risk acceptance criteria we introduced can, hence, apply for various kinds of risk that are prioritized differently by the relevant stakeholders.
In the future, we want to research further how the \emph{budgeting} of risk can be realized and accounted for in the argumentation.
For instance, the specification of tolerable target risk (see~\Cref{subsec:ProductArg}) requires acknowledging that the accepted risk associated with physical harm is influenced by the risk to mobility that society is willing to accept, as parametrization of speed in behavior planning determines the trade-off of physical wellbeing and mobility to all road users.

We also aim to use the specified requirements as a basis for assessing published argumentation approaches to identify potential for improvements.
Complementary to this, we aim to provide in-depth insights into the GSN model underlying this paper.
On the one hand, we thereby want to strengthen the demonstration of the suitability of our requirement-based approach.
On the other hand, we plan to discuss the concrete lines of argumentation against the weaknesses of published argumentations that we identify based on the aforementioned requirement-based evaluation.
Addressing the details of our GSN model will enable us to further delve into some of the challenges highlighted in this article, such as the connection between global and scenario-based risk acceptance criteria, or the use of methods for the quantitative elicitation and propagation of evidence uncertainties.

\section*{ACKNOWLEDGMENT}

We thank Olaf Franke, Jonas Kruss, and Klaus Lamm from MAN Truck \& Bus SE for their substantial contributions to the proposed argumentation framework, as well as Linda Block for her valuable support in improving the language and linguistic style of this paper.
In addition, we highly appreciate the detailed input from the reviewers.
As discussed in the conclusion, we have already been researching open questions, some of which are directly in line with the ideas identified by the reviewers as potentially helpful additions to this paper, and plan to publish corresponding work in the future.


\renewcommand*{\bibfont}{\footnotesize}

\printbibliography


\end{document}